\documentclass[12pt]{article}

\newcommand{\be}{\begin{equation}}
\newcommand{\ee}{\end{equation}}
\newcommand{\ba}{\begin{eqnarray}}
\newcommand{\ea}{\end{eqnarray}}
\newcommand{\ep}{\varepsilon}
\newcommand{\nn}{\nonumber}
\newcommand{\ra}{\rightarrow}
\newcommand{\lra}{\leftrightarrow}

\newcommand{\rom}{\textrm}

\newcommand{\z}{\,\,}
\newcommand{\eqr}[1]{(\ref{#1})}

\begin{document}

\begin{flushright}
{\bf DESY 05--254}\\
{\bf MZ--TH/05--28}\\
\end{flushright}

\vglue 1.4cm
\begin{Large} \begin{bf}
\noindent
Laurent series expansion of a class of massive scalar one-loop integrals 
up to ${\cal O}(\ep^2)$ in terms of multiple polylogarithms.
\end{bf} \end{Large}
\vglue 0.35cm
\parbox{5.4in}
{\leftskip=1.0pc
J.\ G.\ K\"{o}rner$^{\rom a)}$   \\
   {\it  Institut f\"{u}r Physik, Johannes Gutenberg-Universit\"{a}t,
         D-55099 Mainz, Germany}     \\
\\
Z.\ Merebashvili$^{\rom b)}$  \\
   {\it Institute of High Energy Physics and Informatization,
        Tbilisi State University, 0186 Tbilisi, Georgia}   \\
\\
M.\ Rogal$^{\rom c)}$    \\
   {\it Deutsches Elektronen-Synchrotron DESY,
        Platanenallee 6, D-15738 Zeuthen, Germany}     \\
\\
\\
\noindent
In a recent paper we have presented results for a set of massive scalar 
one-loop 
master integrals needed in the NNLO parton model description of the 
hadroproduction of heavy flavors. The one--loop integrals were evaluated in 
$n=4-2\ep$ dimension and the results were presented in terms of a Laurent series 
expansion up to ${\cal O}(\ep^2)$. We found that some of the 
$\ep^2$ coefficients contain 
a new class of functions which we termed the $L$ functions. The $L$ functions are
defined in terms of one--dimensional integrals involving products of logarithm and dilogarithm
functions. In this paper we derive a complete set of algebraic relations that 
allow one to convert the $L$ functions of our previous approach to a sum of 
classical and multiple polylogarithms. Using these results we are now able to 
present the $\ep^2$ coefficients of the one-loop master integrals in terms of 
classical and multiple polylogarithms. 
}

\renewcommand{\thefootnote}{a)}
\footnotetext{Electronic mail: koerner@thep.physik.uni-mainz.de}
\renewcommand{\thefootnote}{b)}
\footnotetext{Electronic mail: zaza@thep.physik.uni-mainz.de}
\renewcommand{\thefootnote}{c)}
\footnotetext{Electronic mail: Mikhail.Rogal@desy.de}

\newpage

               \begin{bf}
\noindent
I. INTRODUCTION
               \end{bf}
\vglue .3cm

Recently, we have calculated the complete set of massive one-loop master integrals 
\cite{KMR} 
needed in the calculation of the next-to-next-to-leading order (NNLO) parton model 
corrections to the hadroproduction of heavy flavors \cite{KMR05}. We used Feynman
parametrization to evaluate the one-loop master integrals in $n=4-2\ep$ dimensions.
We obtained the coefficients of the Laurent 
series expansion of the relevant scalar integrals in terms of the parameter $\ep$ 
up to ${\cal O}(\ep^2)$ as needed for the NNLO calculation. We found that the 
real parts of some of the $\ep^2$ coefficients contain 
a new class of functions which can be written in terms of one--dimensional integral
representations involving products of log and dilog functions. These so--called 
single and triple index $L$ functions cannot be expressed in terms of 
classical 
polylogaritms but can be seen to belong to a generalization of the classical 
polylogarithms which are called multiple polylogarithms. 

Functions analogous to the triple index functions $L_{\sigma_1\sigma_2\sigma_3}$
also arise in the approach of \cite{Andrei} when one analytically
continues their ${\cal O}(\ep^2)$ integral representation for a general vertex 
function. Methods differing from ours have been used for the 
derivation of master $N$-point integrals such as  
the differential equations method \cite{diffeqs} or the nested sum method 
\cite{nested}. Depending on the number of scales involved, 
the results include multiple polylogarithms  \cite{Multilogs} and/or harmonic 
\cite{Remiddi1} or two-dimensional harmonic \cite{Remiddi2} 
polylogarithms. The latter function all are subsets of multiple polylogarithms. 
Presenting our results in terms of multiple polylogarithms will facilitate
a comparison with the results of possible rederivations of the scalar one--loop  
integrals using other methods. It is very likely that future results of 
multiloop calculations will be presented in terms of multiple polylogarithms 
or their subclasses.
Alongside with this the necessary tools will be developed to deal with multiple 
polylogarithms, be it analytically or numerically. In fact, recently a computer code 
has been written for the 
numerical evaluation of the multiple polylogarithms \cite{Vollinga}. It is 
therefore timely that we express the results of \cite{KMR} also in terms of 
multiple polylogarithms.

It is a purpose of this paper to show that the single and triple index 
$L$ functions 
introduced in \cite{KMR} can all be related to multiple polylogarithms. This is
done in explicit form. We are thus able to present our results for the scalar 
massive one--loop master integrals in terms of multiple polylogarithms and 
classical 
polylogarithms \cite{Lewin}. In Sec.~II we recapitulate material on the definition
of the single and triple index $L$ functions as they arise in the approach of
\cite{KMR}. Simple symmetry relations allow one to restrict the discussion to
the triple index $L$ functions $L_{-++}$ and $L_{+++}$, and to the single index
$L$ function $L_{+}$. In Sec.~II we also recapitulate the definition of 
multiple polylogarithms. In the subsequent sections we will 
write down the formulas needed to transform the 
$L$ functions to multiple polylogarithms for general arguments. The general
formulas are not always applicable when the arguments take special values as they
do in the massive one-loop calculation. For these special values one must
carefully 
discuss the limiting behavior of the general formulas. In Sec.~III A 
we derive the 
general formula which relates the $L_{-++}$ functions to the set of multiple 
polylogarithms. 
Section~III B considers special cases of the general relation. 
Similarly, 
Sec.~IV A gives general relations which allow one to express the $L_{+++}$ functions 
in terms of multiple polylogarithms. 
In Sec.~IV B we discuss special cases for the arguments of the 
$L_{+++}$ functions. 
Sections~V A and~V B repeat the discussion for the single 
index $L_{+}$ functions. 
Finally, Sec.~VI presents our conclusions.

As remarked on before, the $L$ functions appear only in the real parts of some
of the ${\cal O}(\ep^2)$ coefficient functions of the masive one--loop
integrals. In the notation of \cite{KMR}
these are the three--point coefficient functions ${\rm Re}\,C_1^{(2)}$, 
${\rm Re}\,C_2^{(2)}$ and ${\rm Re}\,C_5^{(2)}$, and the four--point
coefficient functions ${\rm Re}\,D_1^{(2)}$, ${\rm Re}\,D_2^{(2)}$ and
${\rm Re}\,D_3^{(2)}$. For the sake of brevity we have decided to present
multiple polylogarithm results in this paper only for the four--point coefficient 
function ${\rm Re}\,D_1^{(2)}$. This result is listed in the Appendix.
The corresponding results for the other five coefficient functions  are readily 
available in electronic form \cite{epaps}.

\vglue 1.0cm

               \begin{bf}
\noindent
II. BASIC FEATURES
               \end{bf}
\vglue .3cm

In order to make the paper self--contained, we write down a number of basic 
definitions for the $L$ functions and the multiple polylogarithms in this section, 
as well as some symmetry properties and domains of 
definitions for the single and triple index $L$ functions. These will be of help  
when presenting the 
subsequent material.

The definition for the $L$ functions is as follows \cite{KMR}:
\be
\label{Lfunction}
L_{\sigma_1\sigma_2\sigma_3}(\alpha_1,\alpha_2,\alpha_3,\alpha_4)=
\int_0^1 dy \frac{\ln (\alpha_1+\sigma_1 y) \ln (\alpha_2+\sigma_2 y)
\ln (\alpha_3+\sigma_3 y)}{\alpha_4+y}
\ee
and
\be
\label{Lpfunction}
L_{\sigma_1}(\alpha_1,\alpha_2,\alpha_3,\alpha_4)=
\int_0^1 dy \frac{\ln (\alpha_1+\sigma_1 y) {\rm Li}_2(\alpha_2+\alpha_3 y)
}{\alpha_4+y}.
\ee   
Here the $\sigma_i\,\, (i=1,2,3)$ take the values $\pm 1$ and the $\alpha_j$'s 
are
either integers $\{1,0,-1\}$ or else
kinematical variables. We want to emphasize that the numerical evaluation of 
the $L$ functions is straightforward. 

The $L$ functions possess simple symmetry properties as follows. 
One notices that 
a change of the integration variable $y\rightarrow 1-y$ results
in the identity
\be
L_{\sigma_1} (\alpha_1, \alpha_2, \alpha_3, \alpha_4) = -
L_{-\sigma_1} (\alpha_1 + \sigma_1, \alpha_2 + \alpha_3, -\alpha_3, 
-\alpha_4
- 1)
\ee
which implies that $L_{-}$ can always be related to $L_{+}$, and
vice versa. We have thus written our results for the three-point and 
four-point functions in \cite{KMR} only in terms of the $L_{+}$ 
functions.

Turning to the triple index $L$ function one notices that  $L_{\sigma_1 
\sigma_2 \sigma_3} (\alpha_1, \alpha_2, \alpha_3, \alpha_4)$
is symmetric under permutations of any
two pairs of indices and arguments $\{\sigma_i, \alpha_i\}$ and $\{\sigma_j,
\alpha_j\}$ for $(i\ne j)$. The same change of variables as above 
$y\rightarrow
1-y$ results in
\be
\label{symmetry2}
L_{\sigma_1 \sigma_2 \sigma_3} (\alpha_1, \alpha_2, \alpha_3, \alpha_4) = -
L_{-\sigma_1 -\sigma_2 -\sigma_3} (\alpha_1 + \sigma_1, \alpha_2 + \sigma_2,
\alpha_3 + \sigma_3, -\alpha_4 - 1).
\ee
Therefore, from the eight functions $L_{---}$, $L_{--+}$, $L_{-+-}$,
$L_{+--}$, $L_{-++}$, $L_{+-+}$, $L_{++-}$, and $L_{+++}$ only two are 
independent. We have chosen to write our results in terms of $L_{-++}$ and 
$L_{+++}$.

The domains of definition of the functions $L_{+++}, L_{-++}$, and $L_{+}$ that 
follow from the requirement that these functions take real values can be 
read off from the defining relations Eqs.~(\ref{Lfunction}) and 
(\ref{Lpfunction})
considering the arguments of the log and dilog functions in the integrands, as well 
as from ensuring that the denominator of Eqs.~(\ref{Lfunction}) and 
(\ref{Lpfunction})  does 
not change sign on the integration path. One has
\ba
\label{domain}
 \begin{array}{r@{\quad:\quad}l}
 L_{+++}(\alpha_{1},\alpha_2, \alpha_{3}, \alpha_{4})  &  \alpha_{1}>0
 ,  \alpha_2 > 0, \alpha_{3} > 0, \alpha_{4}<-1 \quad  {\rm or}\quad
 \alpha_{4}>0;  \\
 L_{-++}(\alpha_{1},\alpha_2, \alpha_{3}, \alpha_{4})  &  \alpha_{1} >
 1,  \alpha_2 > 0, \alpha_{3}> 0, \alpha_{4}<-1\quad {\rm or}\quad
 \alpha_{4}>0;  \\
 L_{+}(\alpha_{1},\alpha_2, \alpha_{3}, \alpha_{4})  &  \alpha_{1} >0
 ,  \alpha_2 \leq 1,\alpha_{2}+ \alpha_{3}\leq 1, \alpha_{3}\neq 0,
 \alpha_{4}<-1\quad {\rm or}\quad  \alpha_{4}>0.
  \end{array}
\ea
Looking at the definition of the triple index $L$ function in
(\ref{Lfunction}) one concludes that the boundary points $\alpha_1=0$
and/or $\alpha_2=0$ and/or $\alpha_3=0$ can be included in the domain of
the definition
for $L_{+++}$. The same holds true for  $\alpha_1=1$ and/or $\alpha_2=0$
and/or $\alpha_3=0$ for $L_{-++}$. Also, from the definition of the 
single index
function $L_+$ in (\ref{Lpfunction}) one concludes that
the boundary point $\alpha_1=0$ can be added to its domain of definition.

The points $\alpha_{4}=\{-1, 0\}$ can also be included in the domain if the
values taken by the other parameters $\alpha_{i}$ guarantee the convergence of
the integral. 
We mention that for all of our purposes the conditions (\ref{domain}),
with the boundary points included, are 
satisfied, e.g., our results for the integrals are real.
Nevertheless, it is of course always possible to analytically continue the 
parameters to the complex plane.

There are some further relations for the $L$ functions which result from applying 
integration-by-parts identities. They are not listed here but can be found in 
Appendix~C of \cite{KMR}. They have been used to reduce the set of
$L$ functions occurring in the master integrals to a subset of $L$ functions  
having real values in physical phase space \cite{KMR}.

Multiple polylogarithms are defined as a limit of Z sums \cite{Multilogs}, e.g.,
\ba
\label{zsum} 
Li_{m_{k},...,m_{1}}(x_{k},...,x_{1})=\lim_{n_1 \rightarrow \infty } \sum_{
n_{1} > n_{2}...> n_{k} > 0} 
\frac{x_{1}^{n_{1}}x_{2}^{n_{2}}...x_{k}^{n_{k}}
}{n_{1}^{m_{1}}n_{2}^{m_{2}}...n_{k}^{m_{k}} }.
\ea
The number $w=m_{1}+...+m_{k}$ is called the weight and $k$ is called
the depth of the multiple polylogarithm. The
power series (\ref{zsum}) is convergent for $|x_{i}|<1$, and can be 
analytically
continued via the iterated integral representation:
\ba
\label{intrepr}
Li_{m_{k},...,m_{1}}(x_{k},...,x_{1})=\int \limits_{0}^{x_{1}x_{2}...x_{k}} 
\left(
\frac{dt}{t} \circ \right)^{m_{1}-1}
\frac{dt}{x_{2}x_{3}...x_{k}-t} \circ
\nonumber \\ \left( \frac{dt}{t} \circ \right)^{m_{2}-1}
\frac{dt}{x_{3}...x_{k}-t} \circ ...  \circ  \left( \frac{dt}{t} \circ
\right)^{m_{k}-1}   \frac{dt}{1-t},
\ea
where the following notation is used for the iterated integrals:
\ba
\int \limits_{0}^{\lambda} \frac{dt}{a_{n}-t}\circ ...\circ
\frac{dt}{a_{1}-t} = \int \limits_{0}^{\lambda} \frac{dt_{n}}{a_{n}-t_{n}}
\int \limits_{0}^{t_{n}} \frac{dt_{n-1}}{a_{n-1}-t_{n-1}} \times...\times 
\int
\limits_{0}^{t_{2}}\frac{dt_1}{a_{1}-t_{1}}.
\ea

\vglue 1.0cm 

               \begin{bf}
\noindent
III. TRANSFORMATION OF $L_{-++}$ TO MULTIPLE POLYLOGARITHMS
               \end{bf}
\vglue .3cm

In this section we will show that all our $L_{-++}$ functions can be 
expressed in terms of multiple polylogarithms. 

\vglue 1.0cm
              \begin{bf}
\noindent
A. General case for the $L_{-++}$ function
               \end{bf}
\vglue .3cm

We begin with the $L_{-++}$ function Eq.~(\ref{Lfunction}),
\be
\label{l3gen1}
L_{-++}(\alpha_1,\alpha_2,\alpha_3,\alpha_4)=
\int \limits_0^1 dy \frac{\ln (\alpha_1- y) \ln (\alpha_2+ y)
\ln (\alpha_3 + y)} {\alpha_4 + y} .
\ee
After  changing  the integration variable $y=\alpha_{1} t$ one gets
\ba  
\label{getB27General}
\int \limits_{0}^{1/\alpha_1} dt
\frac{\ln(\alpha_1-\alpha_{1} t) \ln(\alpha_{2}+\alpha_{1}
t)\ln(\alpha_3+\alpha_{1} t)}{\frac{\alpha_{4}}{\alpha_{1}}+t} =
\int \limits_{0}^{1/\alpha_1} dt
\frac{\ln\alpha_1 \ln(\alpha_{2}+\alpha_{1} t)\ln(\alpha_3+\alpha_{1}
t)}{\frac{\alpha_{4}}{\alpha_{1}}+t}\nn\\
+\int \limits_{0}^{1/\alpha_1} dt
\frac{\ln(1-t) [ \ln\alpha_1+\ln(\frac{\alpha_2}{\alpha_1}+t)] [
\ln\alpha_1+\ln(\frac{\alpha_3}{\alpha_1}+ t) ]  
}{\frac{\alpha_4}{\alpha_1}+t}=\nn\\   
\ln\alpha_{1} \int \limits_0^1 dy \frac{ \ln (\alpha_2+ y)\ln (\alpha_3 + 
y)}
{\alpha_4 + y}+\ln^{2}\alpha_1  
\int \limits_{0}^{1/\alpha_1} dt
\frac{\ln(1-t)}{\frac{\alpha_4}{\alpha_1}+t}\nn\\
+\ln\alpha_1
\int \limits_{0}^{1/\alpha_1} dt
\frac{\ln(1-t)\ln(\frac{\alpha_2}{\alpha_1}+ 
t)}{\frac{\alpha_4}{\alpha_1}+t}
+\ln\alpha_1
\int \limits_{0}^{1/\alpha_1} dt
\frac{\ln(1-t)\ln(\frac{\alpha_3}{\alpha_1}+ 
t)}{\frac{\alpha_4}{\alpha_1}+t}~~~\\
+\int \limits_{0}^{1/\alpha_1} dt
\frac{\ln(1-t)\ln(\frac{\alpha_2}{\alpha_1}+ t)
\ln(\frac{\alpha_3}{\alpha_1}+ t) }{\frac{\alpha_4}{\alpha_1}+t}\,
. \nn
 \ea
With the  help of (\ref{intrepr}) the integral in the
second term of the last equation of (\ref{getB27General}) can be  written as \ba
\label{secondForB27} \int \limits_{0}^{1/\alpha_1} dt
\frac{\ln(1-t)}{\frac{\alpha_4}{\alpha_1}+t}= \int
\limits_{0}^{1/\alpha_1}
\frac{dt_1}{-\frac{\alpha_4}{\alpha_1}-t_{1}} \int \limits_{0}^{t_1}
 \frac{dt_2}{1-t_{2}}=Li_{1,1}(-\frac{\alpha_4}{\alpha_1},
-\frac{1}{\alpha_4}  ).
\ea
The third and fourth terms of the last equation in (\ref{getB27General}) contain
integrals of the form
\ba \label{defB23} \int \limits_{0}^{t_m} dt
\frac{ \ln(1-t) \ln(\beta_1 + t)} {\beta_2 + t}\, . \ea
To express  
such integrals in terms of multiple polylogarithms one proceeds as
follows:
\ba \label{getB23} -Li_{1,1,1} \left(
-\beta_2,\frac{\beta_1}{\beta_2},\frac{-t_m}{\beta_1} \right)=
 \int \limits_{0}^{t_m}  \frac{dt_{2}}{\beta_1 + t_{2}} \int 
\limits_{0}^{t_{2}}
dt_{1} \frac{\ln(1-t_{1})} {\beta_2 + t_{1}}=
 \int \limits_{0}^{t_m}  dt_{1} \frac{\ln(1-t_{1})} {\beta_2 + t_{1}}  \int
\limits_{t_{1}}^{t_m} \frac{dt_{2}}{\beta_1 + t_{2}}=       \nn      \\
\ln(\beta_1 + t_m)  \int \limits_{0}^{t_m} dt_{1} \frac{ \ln(1-t_{1})} 
{\beta_2
+ t_{1}}
- \int \limits_{0}^{t_m} dt_{1} \frac{ \ln(1-t_{1}) \ln(\beta_1 + t_{1})}
{\beta_2 + t_{1}}  =                ~~~                \\
\nn
\ln(\beta_1 + t_m) Li_{1,1} \left( -\beta_2, \frac{-t_m}{\beta_2} \right)
- \int \limits_{0}^{t_m} dt_{1} \frac{ \ln(1-t_{1})\ln(\beta_1 + t_{1})}
{\beta_2 + t_{1}}\, . 
\ea
In the first line of (\ref{getB23}) we have changed the order of integration in the 
two-dimensional integral.
We shall frequently use this trick further on. 
From Eq.~(\ref{getB23})  one immediately concludes that
\be
\label{B23}
 \int \limits_{0}^{t_m} dt \frac{ \ln(1-t)
\ln(\beta_1 + t)} {\beta_2 + t}=
Li_{1,1,1} \left( -\beta_2,\frac{\beta_1}{\beta_2},\frac{-t_m}{\beta_1}
\right) +
\ln\left(\beta_1 + t_m\right) Li_{1,1} \left( -\beta_2, \frac{-t_m}{\beta_2}
\right).
\ee
Let us now turn to the more involved integral [first term of 
Eq.~(\ref{getB27General})]:
\ba
\label{getB23A}
\int \limits_0^1 dy \frac{ \ln (\alpha_2+ y)\ln (\alpha_3 + y)} {\alpha_4 + 
y}\,
\stackrel{y\rightarrow-\alpha_2 t }{\, =\, } \,
-\int \limits_0^{-1/\alpha_2} dt \frac{[\ln\alpha_2+  \ln (1- t)] \ln 
(\alpha_3
-\alpha_2 t)} {\frac{\alpha_4}{\alpha_2} -t}=\nn\\
-\ln\alpha_2\int \limits_0^{-1/\alpha_2} dt \frac{\ln (\alpha_3 -\alpha_2 
t)}
{\frac{\alpha_4}{\alpha_2} -t}
-\int \limits_0^{-1/\alpha_2} dt \frac{\ln (1- t)[\ln\alpha_2 + \ln
(\frac{\alpha_3}{\alpha_2} -t)]} {\frac{\alpha_4}{\alpha_2} -t}=\\
+\ln\alpha_2\int \limits_0^{1} dy \frac{\ln (\alpha_3 +y)} {\alpha_4 +y}
-\ln\alpha_2 \int \limits_0^{-1/\alpha_2} dt \frac{\ln (1- t)}
{\frac{\alpha_4}{\alpha_2} -t}
-\int \limits_0^{-1/\alpha_2} dt \frac{\ln (1- t)\ln 
(\frac{\alpha_3}{\alpha_2}
-t)} {\frac{\alpha_4}{\alpha_2} -t}\nn \, .
\ea
The integral in the first term can be expressed as
\ba
\label{B28}
\int \limits_0^{1} dy \frac{\ln (\alpha_3 +y)} {\alpha_4 +y}\,
\stackrel{y\rightarrow-\alpha_3 t }{\, =\, } \,
-\int \limits_0^{-1/\alpha_3} dt \frac{[\ln\alpha_3 +\ln (1 -t)]}
{\frac{\alpha_4}{\alpha_3} -t}=\nn\\
\ln\alpha_3\ln\left(\frac{\alpha_{4}+1}{\alpha_{4}}\right)+Li_{1,1}
\left(\frac{\alpha_4}{ \alpha_3 },-\frac{1}{ \alpha_3 } \right  )\, .
\ea
The integral in the second term can be written as 
\ba
\label{B28A}
\int \limits_0^{-1/\alpha_2} dt \frac{\ln (1- t)} {\frac{\alpha_4}{\alpha_2}
-t}=-Li_{1,1}\left( \frac{\alpha_4}{ \alpha_2},-\frac{1}{ \alpha_2 }   
\right)\, .
\ea
The third term from the last line of Eq.~(\ref{getB23A}) has a form which is  
an
analogue of the integral (\ref{defB23})
and can be calculated in a similar way,
\ba
\label{B23C}
\int \limits_{0}^{t_m} dt \frac{ \ln(1-t)
\ln(\beta_1 - t)} {\beta_2 - t}= Li_{1,1,1} \left(
\beta_2,\frac{\beta_1}{\beta_2},\frac{t_m}{\beta_1}
\right) +
\ln\left(\beta_1 - t_m\right) Li_{1,1} \left( \beta_2, \frac{t_m}{\beta_2}
\right)\, .
\ea
Combining the Eqs.~(\ref{B28}),~(\ref{B28A}) and~(\ref{B23C}) we arrive at 
the
result for Eq.~(\ref{getB23A}),
\ba
\label{B23A}
\int \limits_0^1 dy \frac{ \ln (\alpha_2+ y)\ln (\alpha_3 + y)} {\alpha_4 + 
y}=
Li_{1,1,1}\left(\frac {\alpha_4}{\alpha_2},
\frac{\alpha_{3}}{\alpha_{4}},-\frac{1}{\alpha_{3}}\right)
+\ln\alpha_{2}Li_{1,1}\left( 
\frac{\alpha_{4}}{\alpha_{3}},-\frac{1}{\alpha_{4}}
\right )\nn \\
+\ln(1+\alpha_{3})Li_{1,1}\left(
\frac{\alpha_{4}}{\alpha_{2}},-\frac{1}{\alpha_{4}} \right )
+\ln\alpha_{2}\ln\alpha_{3}\ln\left(\frac{\alpha_{4}+1}{\alpha_{4}}\right).
\ea
Because the initial integrand is symmetric under the exchange of the 
parameters
$\alpha_{2}$ and $\alpha_{3}$,
the rhs of (\ref{B23A}) can be rewritten in a symmetric form if desired.

We are now left with the fifth term in (\ref{getB27General}). The fifth term 
is an integral of the type
\ba
\label{typeB27}
\int \limits_{0}^{t_m} dt \frac{ \ln(1-t)    \ln(\gamma_1 + t)
\ln(\gamma_2 + t)} {\gamma_3 +t}\, .
\ea
In order to express such integrals in terms of multiple polylogarithms one can 
perform the
following chain of transformations resulting in a multiple polylogarithm of weight four:
\ba
\label{l3gen4}
- Li_{1,1,1,1}\left(-\gamma_3, \frac{\gamma_2}{\gamma_3},   
\frac{\gamma_1}{\gamma_2}, \frac{-t_m}{\gamma_1} \right) =
 \int \limits_{0}^{t_m} \frac{dt_{4}}{\gamma_1 +t_{4}} \int 
\limits_{0}^{t_{4}}
\frac{dt_{3}}{\gamma_2 +t_{3}}
 \int \limits_{0}^{t_{3}}  \frac{dt_{2}}{\gamma_3 +t_{2}} \int 
\limits_{0}^{t_{2}}
\frac{dt_{1}}{1-t_{1}}   =             \\
\nn
- \int \limits_{0}^{t_m} \frac{dt_{4}}{\gamma_1 +t_{4}} \int 
\limits_{0}^{t_{4}}
\frac{dt_{3}}{\gamma_2 +t_{3}}
  \int \limits_{0}^{t_{3}} d t_{2} \frac{\ln(1-t_{2})}{\gamma_3 +t_{2}}  =
- \int \limits_{0}^{t_m} \frac{dt_{4}}{\gamma_1 +t_{4}} \int 
\limits_{0}^{t_{4}}
d t_{2} \frac{\ln(1-t_{2})}{\gamma_3 +t_{2}}
 \int \limits_{t_{2}}^{t_{4}} \frac{dt_{3}}{\gamma_2 +t_{3}}=       \\
\nn
- \int \limits_{0}^{t_m} dt_{4} \frac{\ln(\gamma_2 +t_{4})}{\gamma_1 +t_{4}}  
\int
\limits_{0}^{t_{4}} d t_{2} \frac{ \ln(1-t_{2})}{\gamma_3 + t_{2}} +
 \int \limits_{0}^{t_m} \frac{dt_{4}}{\gamma_1 + t_{4}} \int 
\limits_{0}^{t_{4}}
dt_{2} \frac{\ln(1-t_{2}) \ln(\gamma_2 + t_{2})}{\gamma_3 
+t_{2}}=\label{big}  \\
\nn
-I'(t_m) + \int \limits_{0}^{t_m} dt_{2}  \frac{ \ln(1-t_{2}) \ln(\gamma_2 +
t_{2})} {\gamma_3 +t_{2}}  \int \limits_{t_{2}}^{t_m}  
\frac{dt_{4}}{\gamma_1
 + t_{4}}=                           \\
\nn
-I'(t_m) + I''(t_m)
 - \int \limits_{0}^{t_m} dt_{2} \frac{ \ln(\gamma_1 + t_{2})
\ln(\gamma_2 + t_{2})\ln(1-t_{2})} {\gamma_3 +t_{2}},
\ea
where we have introduced the notation
\ba
\label{iprimes}
I'(t_m) =  \int \limits_{0}^{t_m} dt_{4} \frac{\ln(\gamma_2 
+t_{4})}{\gamma_1
+t_{4}}  \int \limits_{0}^{t_{4}} d t_{2} \frac{ \ln(1-t_{2})}{\gamma_3 + 
t_{2}},
\\    \nn
I''(t_m) = \ln(\gamma_1 + t_m) \int \limits_{0}^{t_m} dt_{2} \frac{ 
\ln(1-t_{2})
\ln(\gamma_2 + t_{2})} {\gamma_3 + t_{2}}.
\ea
The third term on the last line of (\ref{l3gen4}) is exactly the integral of
the required type Eq.~(\ref{typeB27}).

The integral in $I''(t_m)$ has the form of  (\ref{B23}). For the integral
$I'(t_m) $ we write
\be
I'(t_m) =  \int \limits_{0}^{t_m} dt_{4} \frac{\ln(\gamma_2 
+t_{4})}{\gamma_1
+t_{4}}  \int \limits_{0}^{t_{4}} d t_{2} \frac{ \ln(1-t_{2})}{\gamma_3 + 
t_{2}} =
\int \limits_{0}^{t_m} dt_{4} \frac{\ln(\gamma_2 + t_{4})} {\gamma_1 + 
t_{4}}
Li_{1,1}\left(-\gamma_3,\frac{-t_{4}}{\gamma_3}\right) .
\ee
On the other, hand one has
\ba
Li_{1,1,1,1} \left( -\gamma_3, \frac{\gamma_1}{\gamma_3}, \frac{\gamma_2}
{\gamma_1}, \frac{-t_m}{\gamma_2} \right) =
\int \limits_{0}^{t_m} \frac{ dt_{2}}{\gamma_2 + t_{2}}  
\int \limits_{0}^{t_{2}} \frac{dt_{1}} {\gamma_1 + t_{1}}
Li_{1,1} \left( -\gamma_3, \frac{-t_{1}}{\gamma_3} \right)=        \\
\nn
 \int \limits_{0}^{t_m} \frac{dt_{1}}{\gamma_1 + t_{1}}   
Li_{1,1} \left( -\gamma_3, \frac{-t_{1}}{\gamma_3} \right)
 \int \limits_{t_{1}}^{t_m}  \frac{ dt_{2}}{\gamma_2
+ t_{2}}=                                  \\
\nn
\ln(\gamma_2 + t_m) \int \limits_{0}^{t_m} \frac{dt_{1}} {\gamma_1 + t_{1}}
Li_{1,1} \left( -\gamma_3, \frac{-t_{1}}{\gamma_3} \right)
- \int \limits_{0}^{t_m} dt_{1} \frac{ \ln(\gamma_2 + t_{1})}{\gamma_1 + 
t_{1}}
Li_{1,1}\left( -\gamma_3, \frac{-t_{1}}{\gamma_3} \right) =               \\
\nn
- \ln(\gamma_2 + t_m) Li_{1,1,1}\left( -\gamma_3, \frac{\gamma_1}{\gamma_3},
\frac{-t_m}{\gamma_1} \right) - I'(t_m).
\ea
One then concludes that  
\be
\label{sub2}
I'(t_m) = - Li_{1,1,1,1} \left( -\gamma_3, \frac{\gamma_1}{\gamma_3},
\frac{\gamma_2} {\gamma_1}, \frac{-t_m}{\gamma_2} \right) - \ln(\gamma_2 + 
t_m)
Li_{1,1,1} \left( -\gamma_3, \frac{\gamma_1}{\gamma_3}, 
\frac{-t_m}{\gamma_1}
\right).
\ee
Finally, substituting $I'(t_m) $ and $I''(t_m)$  into Eq.~(\ref{l3gen4})
we write down the result for the integral of the required type 
Eq.~(\ref{typeB27}),
\ba
\label{B27}
\int \limits_{0}^{t_m} dt \frac{  \ln(1 - t)   \ln(\gamma_1 + t) 
\ln(\gamma_2 + t)}
 {\gamma_3 + t} =
\ln(\gamma_1 + t_m) \ln(\gamma_2 + t_m)
Li_{1,1}\left( -\gamma_3, \frac{-t_m} {\gamma_3} \right) + \nn
\\
  \ln(\gamma_2 + t_m) Li_{1,1,1}\left( -\gamma_3, \frac{\gamma_1} 
{\gamma_3},
\frac{-t_m}{\gamma_1} \right)
+ \ln(\gamma_1 + t_m)
Li_{1,1,1}\left( -\gamma_3, \frac{\gamma_2} {\gamma_3}, \frac{-t_m}
{\gamma_2} \right)                              \\
+ Li_{1,1,1,1}\left( -\gamma_3, \frac{\gamma_2} {\gamma_3}, \frac{\gamma_1}
{\gamma_2}, \frac{-t_m}{\gamma_1} \right)
+ Li_{1,1,1,1} \left( -\gamma_3, \frac{\gamma_1} {\gamma_3}, \frac{\gamma_2}
{\gamma_1}, \frac{-t_m}{\gamma_2} \right). \nn
\ea

We are now in the position to collect all required contributions to express 
the $L_{-++}$ function in terms of multiple polylogarithms.
Taking into account Eqs.~(\ref{secondForB27}), (\ref{B23}), (\ref{B23A}), and
(\ref{B27}) we obtain
\ba
\label{B27General}
L_{-++}(\alpha_1,\alpha_2,\alpha_3,\alpha_4)=
Li_{1,1,1,1}\left(-\frac{\alpha_4}{\alpha_1},  \frac{\alpha_2}{\alpha_4},
\frac{\alpha_3}{\alpha_2},-  \frac{1}{\alpha_3}\right)\nn\\   
+Li_{1,1,1,1}\left(-\frac{\alpha_4}{\alpha_1},  \frac{\alpha_3}{\alpha_4},
\frac{\alpha_2}{\alpha_3}, - \frac{1}{\alpha_2}\right)
+\ln\alpha_{1} Li_{1,1,1}\left(\frac{\alpha_4}{\alpha_2},
\frac{\alpha_3}{\alpha_4}, - \frac{1}{\alpha_3}\right)\nn\\
+\ln(1+\alpha_{2}) Li_{1,1,1}\left(-\frac{\alpha_4}{\alpha_1},
\frac{\alpha_3}{\alpha_4}, - \frac{1}{\alpha_3}\right)
+\ln(1+\alpha_{3}) Li_{1,1,1}\left(-\frac{\alpha_4}{\alpha_1},
\frac{\alpha_2}{\alpha_4}, - \frac{1}{\alpha_2}\right)\nn\\
+\ln\alpha_{1}\ln\alpha_{2} Li_{1,1}\left(\frac{\alpha_4}{\alpha_3}, -
\frac{1}{\alpha_4}\right)
+\ln\alpha_{1}\ln(1+\alpha_{3}) Li_{1,1}\left(\frac{\alpha_4}{\alpha_2}, -
\frac{1}{\alpha_4}\right) \\
+\ln(1+\alpha_{2})\ln(1+\alpha_{3}) 
Li_{1,1}\left(-\frac{\alpha_4}{\alpha_1}, -
\frac{1}{\alpha_4}\right)
+\ln\alpha_{1}\ln\alpha_{2}\ln\alpha_{3}
\ln\left(\frac{\alpha_{4}+1}{\alpha_{4}}\right).\nn
\ea
Some remarks are in order at this place. The final formula~(\ref{B27General}) 
contains multiple polylogarithms up to weight four.
All multiple polylogarithms up to weight three can be expressed in terms of 
logarithms and clasical polylogarithms $Li_2$ and $Li_3$ .
This fact is used by us when we reexpress our results for the massive
scalar integrals in terms of multiple polylogarithms, i.e., our final results 
will contain only multiple polylogarithms of weight four.
For the variables $\alpha_{i}$ the conditions~(\ref{domain}) are assumed.
But in the results for the massive scalar integrals there are also cases
when $\alpha_{1}=1$ and/or $ \alpha_{2}=0$ and/or $ \alpha_{3}=0$ and/or
$\alpha_{4}=\{-1,0\}$.
In such cases the general formula ~(\ref{B27General}) is no longer valid and
these cases must be studied separately.

\vglue 1.0cm
              \begin{bf}
\noindent   
B. Special cases for the $L_{-++}$ function
               \end{bf}
\vglue .3cm

In the Laurent series expansion of the massive scalar one-loop integrals one
encounters special values of the arguments $\alpha_{i}$ for which the general
formula Eq.~(\ref{B27General}) no longer applies. This is quite obvious from
the list of special cases discussed in the following. 

\vglue 1.0cm
              \begin{bf}
\noindent
{\em 1. ${\bf \alpha_{1}=1, \z \alpha_{4}=0}$ }
               \end{bf}
\vglue .3cm

In such case one can make use of Eq.~(\ref{B27}).
One should find the limit of the expression on the  right-hand side for $t_{m}=1,
\gamma_{3}\rightarrow 0$. One obtains
\ba
\label{getChangeLmpp}
 \int \limits_{0}^{1} dt \frac{  \ln(1 - t)   \ln(\gamma_1 + t) \ln(\gamma_2 + t)}
 {t} =\hspace{9.8cm} \nn\\
\lim_{\gamma_{3} \rightarrow 0} \Big\{ \ln(\gamma_1 + 1) \ln(\gamma_2 + 1)
\int \limits_{0}^{1} \frac{dt_{2}}{-\gamma_3 -t_{2}} \int
\limits_{0}^{t_{2}}\frac{dt_{1}}{1 -t_{1}}
+
\ln(\gamma_2 + 1)
\int \limits_{0}^{1} \frac{dt_{3}}{-\gamma_1 -t_{3}}\int \limits_{0}^{t_{3}}
\frac{dt_{2}}{-\gamma_3 -t_{2}} \int \limits_{0}^{t_{2}}\frac{dt_{1}}{1 -t_{1}} \nn\\
+ \ln(\gamma_1 + 1)
\int\limits_{0}^{1} \frac{dt_{3}}{-\gamma_2 -t_{3}}\int \limits_{0}^{t_{3}}   
\frac{dt_{2}}{-\gamma_3 -t_{2}} \int \limits_{0}^{t_{2}}\frac{dt_{1}}{1 -t_{1}}
+ \int \limits_{0}^{1} \frac{dt_{4}}{-\gamma_1 -t_{3}}\int \limits_{0}^{t_{4}}
\frac{dt_{3}}{-\gamma_2 -t_{3}}
\int \limits_{0}^{t_{3}} \frac{dt_{2}}{-\gamma_3 -t_{2}} \int
\limits_{0}^{t_{2}}\frac{dt_{1}}{1 -t_{1}} \nn \\
+ \int \limits_{0}^{1} \frac{dt_{4}}{-\gamma_2 -t_{3}}\int \limits_{0}^{t_{4}}
\frac{dt_{3}}{-\gamma_1 -t_{3}}
\int \limits_{0}^{t_{3}} \frac{dt_{2}}{-\gamma_3 -t_{2}} \int
\limits_{0}^{t_{2}}\frac{dt_{1}}{1 -t_{1}}
  \Big \}=
- \ln(\gamma_1 + 1) \ln(\gamma_2 + 1)
\int \limits_{0}^{1} \frac{dt_{2}}{t_{2}} \int \limits_{0}^{t_{2}}\frac{dt_{1}}{1
-t_{1}}\nn\\
-\ln(\gamma_2 + 1)
\int \limits_{0}^{1} \frac{dt_{3}}{-\gamma_1 -t_{3}}\int \limits_{0}^{t_{3}}
\frac{dt_{2}}{t_{2}} \int \limits_{0}^{t_{2}}\frac{dt_{1}}{1 -t_{1}}
- \ln(\gamma_1 + 1)
\int\limits_{0}^{1} \frac{dt_{3}}{-\gamma_2 -t_{3}}\int \limits_{0}^{t_{3}}
\frac{dt_{2}}{t_{2}} \int \limits_{0}^{t_{2}}\frac{dt_{1}}{1 -t_{1}} \nn\\
-\int \limits_{0}^{1} \frac{dt_{4}}{-\gamma_1 -t_{3}}\int \limits_{0}^{t_{4}}
\frac{dt_{3}}{-\gamma_2 -t_{3}}
\int \limits_{0}^{t_{3}} \frac{dt_{2}}{t_{2}} \int
\limits_{0}^{t_{2}}\frac{dt_{1}}{1 -t_{1}}
- \int \limits_{0}^{1} \frac{dt_{4}}{-\gamma_2 -t_{3}}\int \limits_{0}^{t_{4}}
\frac{dt_{3}}{-\gamma_1 -t_{3}}
\int \limits_{0}^{t_{3}} \frac{dt_{2}}{t_{2}} \int
\limits_{0}^{t_{2}}\frac{dt_{1}}{1 -t_{1}}\, .\nn\\
\ea
In order to get the expression under the sign of the limit in 
Eq.~(\ref{getChangeLmpp})
one applies the definition (\ref{intrepr}) for the multiple polylogarithms in
Eq.~(\ref{B27}).
Using the same definition for the final multidimensional integrals in
(\ref{getChangeLmpp})
and making the change $\gamma_{1}\rightarrow\alpha_{2},
\gamma_{2}\rightarrow\alpha_{3} $ one finally arrives at the result
for the case $\alpha_{1}=1$ and $\alpha_{4}=0$,
\ba
\label{ChangeLmpp}
L_{-++}\left(1,\alpha_{2},\alpha_{3},0 \right)=
-Li_{2,1,1}\left(-\alpha_{3},\frac{\alpha_{2}}{\alpha_{3}}, -\frac{1}{\alpha_{2}}
\right)
-Li_{2,1,1}\left(-\alpha_{2},\frac{\alpha_{3}}{\alpha_{2}}, -\frac{1}{\alpha_{3}}
\right) .\nn\\
-\ln(\alpha_{3} + 1)Li_{2,1}\left(-\alpha_{2},-\frac{1}{\alpha_{2}}   \right)
-\ln(\alpha_{2} + 1)Li_{2,1}\left(-\alpha_{3},-\frac{1}{\alpha_{3}}   \right)\nn\\
-\ln(\alpha_{2} + 1) \ln(\alpha_{3} + 1)\zeta(2) .
\ea

\vglue 1.0cm
              \begin{bf}
\noindent
{\em 2. ${\bf \alpha_{1}=1,  \z  \alpha_{2}=\alpha_{3}=0}$ }
               \end{bf}
\vglue .3cm

For these values of the parameters $\alpha_{i}$ one has an integral of the very
simple form
\ba
L_{-++}(1,0,0,\alpha_{4})=\int \limits_0^1 dy \frac{\ln (1- y) \ln^{2}y} {\alpha_4 +
y}\, . \nn
\ea
After  a change of  variable $y\rightarrow 1-t$ one gets
\ba
\int \limits_0^1 dt \frac{\ln t \ln^{2}(1-t)} {\alpha_4+1 - t}=- \int \limits_0^1
dt_{1} \frac{ \ln^{2}(1-t_{1})} {\alpha_4+1 - t_{1}}
\int\limits_{t_{1}}^{1} \frac{dt_{2}}{t_{2}}=\nn\\
- \int \limits_0^1 \frac{dt_{2}}{t_{2}} \int\limits_{0}^{t_{2}} dt_{1}\frac{
\ln^{2}(1-t_{1})} {\alpha_4+1 - t_{1}}=
-2 \int \limits_0^1  \frac{dt_{2}}{t_{2}} \int\limits_{0}^{t_{2}} \frac{dt_{1}}
{\alpha_4+1 - t_{1}} \int \limits_0^{t_{1}} \frac{dt_{3}} {1 - t_{3}}
\int \limits_0^{t_{3}} \frac{dt_{4}} {1 - t_{4}}\, .
\ea
Applying the definition~(\ref{intrepr}) one obtains
\ba
L_{-++}(1,0,0,\alpha_{4})=-2 Li_{1,1,2}\left(1,\alpha_{4}+1,\frac{1}{\alpha_{4}+1}
\right).
\ea

\vglue 1.0cm
              \begin{bf}
\noindent
{\em 3. ${\bf \alpha_{1}=1,  \z  \alpha_{2}=0}$  (and ${\bf \alpha_{4}=-1}$) 
}
               \end{bf}
\vglue .3cm

We shall again find the limit of the rhs of (\ref{B27}) for $t_{m}=1$ and $
\gamma_{1}\rightarrow 0$.
The first and the third terms are equal to 0 because of the limit $
\lim_{\gamma_1\rightarrow 0 }   \ln(\gamma_{1}+1)=0$. The  other terms
transform into
\ba
\lim_{\gamma_1\rightarrow 0 } Li_{1,1,1}\left( -\gamma_3, \frac{\gamma_1} {\gamma_3},
\frac{-1}{\gamma_1} \right) = -Li_{1,2}\left(-\gamma_{3},-\frac{1} {\gamma_3}
\right), \nn \\
\lim_{\gamma_{1}\rightarrow 0 }
Li_{1,1,1,1}\left( -\gamma_3, \frac{\gamma_2} {\gamma_3}, \frac{\gamma_1}{\gamma_2},
\frac{-t_m}{\gamma_1} \right)  =
-Li_{1,1,2}\left( -\gamma_3, \frac{\gamma_2} {\gamma_3}, \frac{-1}{\gamma_2}
\right),\nn\\
\lim_{\gamma_{1}\rightarrow 0 }
 Li_{1,1,1,1} \left( -\gamma_3, \frac{\gamma_1} {\gamma_3}, \frac{\gamma_2}
{\gamma_1}, \frac{-t_m}{\gamma_2} \right)=-Li_{1,2,1}\left( -\gamma_3,
\frac{\gamma_2}{\gamma_3}, \frac{-1}{\gamma_2} \right). \nn
\ea
Finally we write
\ba
L_{-++}\left(1, 0, \alpha_{3}, \alpha_{4} \right)=-Li_{1,1,2}\left( -\alpha_4,
\frac{\alpha_3} {\alpha_4}, \frac{-1}{\alpha_3} \right)
-Li_{1,2,1}\left( -\alpha_4, \frac{\alpha_3} {\alpha_4}, \frac{-1}{\alpha_3}
\right)\nn\\
-\ln(\alpha_{3}+1)Li_{1,2}\left(-\alpha_{4},-\frac{1} {\alpha_4}  \right).
\ea
For the special case $\alpha_{4}=-1$ one gets
\ba
L_{-++}\left(1, 0, \alpha_{3}, -1 \right)=-Li_{1,1,2}\left( 1, -\alpha_3,
\frac{-1}{\alpha_3} \right)
-Li_{1,2,1}\left( 1,- \alpha_3, \frac{-1}{\alpha_3} \right)\nn\\
-\ln(\alpha_{3}+1)\, \zeta(3).
\ea

\vglue 1.0cm
              \begin{bf}
\noindent
{\em 4. ${\bf \alpha_{2}=\alpha_{3}=0}$ \z (and ${\bf \alpha_{4}=-1}$) }
               \end{bf}
\vglue .3cm

In  this case one proceeds along the following lines:
\ba
L_{-++}(\alpha_1, 0, 0,  \alpha_4)=
\int \limits_0^1 dy \frac{\ln (\alpha_1- y) \ln^{2} y} {\alpha_4 + y} \,
\stackrel{y\rightarrow1-t }{\, =\, } \,
\int \limits_0^1 dt \frac{\ln (\alpha_1- 1+t) \ln^{2}(1-t)} {\alpha_4 +1- t}=\nn\\
-\int \limits_0^1 dt_{1} \frac{\ln^{2}(1-t_{1})} {-\alpha_4 -1+ t_{1}}\int
\limits_{-\alpha_{1}+2}^{t_{1}}\frac{dt_{2}} {\alpha_1 -1+ t_{2}}=
-\int \limits_0^1 dt_{1} \frac{\ln^{2}(1-t_{1})} {-\alpha_4 -1+ t_{1}}
\left \{  \int \limits_{1}^{t_{1}} +   \int \limits_{-\alpha_{1}+2}^{1} \right \}
\frac{dt_{2}} {\alpha_1 -1+ t_{2}}=\nn\\
\int \limits_0^1  \frac{dt_{2}} {\alpha_1 -1+ t_{2}} \int \limits_0^{t_{2}}dt_{1}
\frac{\ln^{2}(1-t_{1})} {-\alpha_4 -1+ t_{1}}
-\ln\alpha_{1} \int \limits_0^1 dt_{1} \frac{\ln^{2}(1-t_{1})} {-\alpha_4 -1+
t_{1}}= \nn\\
2 \int \limits_0^1  \frac{dt_{2}} {\alpha_1 -1+ t_{2}} \int \limits_0^{t_{2}}
\frac{dt_{1}} {-\alpha_4 -1+ t_{1}}
\int \limits_0^{t_{1}}\frac{dt_{3}}{1-t_{3}}\int 
\limits_0^{t_{3}}\frac{dt_{4}}{1-t_{4}}
-2\ln\alpha_{1} {\rm Li}_{3}\left(-\frac{1}{\alpha_{4}} \right) .\nn\\
\ea
Using the definition (\ref{intrepr}) we arrive at the result
\ba
L_{-++}(\alpha_1, 0, 0,  \alpha_4)
=2 Li_{1,1,1,1}\left(1, \alpha_{4}+1, \frac{1-\alpha_{1}}{\alpha_{4}+1},
\frac{1}{1-\alpha_{1}} \right)-2\ln\alpha_{1} {\rm
Li}_{3}\left(-\frac{1}{\alpha_{4}} \right)\, .
\ea
For the case $\alpha_{4}=-1$ one obtains
\ba
L_{-++}(\alpha_1, 0, 0, -1)=-2 Li_{1,2,1}\left(1, 1- \alpha_{1},
\frac{1}{1-\alpha_{1}} \right)-2\ln\alpha_{1}\, \zeta(3)\, .
\ea

\vglue 1.0cm 
              \begin{bf}
\noindent
{\em 5. ${\bf \alpha_{2}=0}$  (and ${\bf \alpha_{4}=-1}$) }
               \end{bf}
\vglue .3cm

For this integral we change the integration variable $y \rightarrow 1-t $,
\ba
\label{changeB27A}
\int \limits_0^1 dy \frac{\ln (\alpha_1- y) \ln y  \ln(\alpha_{3}+y)} {\alpha_4 + y}=
\int \limits_0^1 dt \frac{\ln (1- t)  \ln(\alpha_{1}-1+t) \ln(\alpha_{3}+1-t)}
{\alpha_4+1 - t}= \nn \\
\int \limits_0^1 dt \frac{\ln (1- t)  \ln(\gamma_{1}+t) \ln(\gamma_{2}-t)}
{\gamma_{3} - t} \, .
\ea
One notes that the last integral is an analogue of the integral  in Eq.~(\ref{B27}).
The calculation proceeds in a similar way,
\ba  
\label{getB27A}
- Li_{1,1,1,1}\left(\gamma_3, \frac{\gamma_2}{\gamma_3},
-\frac{\gamma_1}{\gamma_2},- \frac{1}{\gamma_1} \right) =
 \int \limits_{0}^{1} \frac{dt_{4}}{\gamma_1 +t_{4}} \int \limits_{0}^{t_{4}}
\frac{dt_{3}}{\gamma_2 -t_{3}}
 \int \limits_{0}^{t_{3}}  \frac{dt_{2}}{\gamma_3 -t_{2}} \int \limits_{0}^{t_{2}}
\frac{dt_{1}}{1-t_{1}}   =       \nn      \\
\nn
- \int \limits_{0}^{t_m} \frac{dt_{4}}{\gamma_1 +t_{4}} \int \limits_{0}^{t_{4}}
\frac{dt_{3}}{\gamma_2 -t_{3}}
  \int \limits_{0}^{t_{3}} d t_{2} \frac{\ln(1-t_{2})}{\gamma_3 -t_{2}}  =
- \int \limits_{0}^{t_m} \frac{dt_{4}}{\gamma_1 +t_{4}} \int \limits_{0}^{t_{4}}
d t_{2} \frac{\ln(1-t_{2})}{\gamma_3 -t_{2}}
 \int \limits_{t_{2}}^{t_{4}} \frac{dt_{3}}{\gamma_2 -t_{3}}=   \\
\nn
\int \limits_{0}^{1} dt_{4} \frac{\ln(\gamma_2 -t_{4})}{\gamma_1 +t_{4}}  \int  
\limits_{0}^{t_{4}} d t_{2} \frac{ \ln(1-t_{2})}{\gamma_3 -t_{2}} -
 \int \limits_{0}^{1} \frac{dt_{4}}{\gamma_1 + t_{4}} \int \limits_{0}^{t_{4}}
dt_{2} \frac{\ln(1-t_{2}) \ln(\gamma_2 - t_{2})}{\gamma_3 - t_{2}}= \\
\nn
Y'(1) - \int \limits_{0}^{1} dt_{2}  \frac{ \ln(1-t_{2}) \ln(\gamma_2 -
t_{2})} {\gamma_3 -t_{2}}  \int \limits_{t_{2}}^{1}  \frac{dt_{4}}{\gamma_1
 + t_{4}}=      \nn\\
Y'(1)
-\ln(\gamma_{1}+1) \int \limits_{0}^{1} dt_{2}  \frac{ \ln(1-t_{2}) \ln(\gamma_2
-t_{2})} {\gamma_3 -t_{2}}+
  \int \limits_{0}^{1} dt_{2} \frac{ \ln(1-t_{2})   \ln(\gamma_1 + t_{2})
\ln(\gamma_2 - t_{2})} {\gamma_3 -t_{2}}=\nn\\
=Y'(1)
-Y''(1)+
  \int \limits_{0}^{1} dt_{2} \frac{ \ln(1-t_{2})   \ln(\gamma_1 + t_{2})
\ln(\gamma_2 - t_{2})} {\gamma_3 -t_{2}}\, , \nn\\
\ea
where we have introduced the notation
\ba
\label{yprimes}
Y'(t_m) =  \int \limits_{0}^{t_m} dt_{4} \frac{\ln(\gamma_2 - t_{4})}{\gamma_1
+t_{4}}  \int \limits_{0}^{t_{4}} d t_{2} \frac{ \ln(1-t_{2})}{\gamma_3 - t_{2}},\nn
\\
Y''(t_m) = \ln(\gamma_1 + t_m) \int \limits_{0}^{t_m} dt_{2} \frac{ \ln(1-t_{2})
\ln(\gamma_2 - t_{2})} {\gamma_3 - t_{2}}.
\ea
The last term in (\ref{getB27A}) is the required integral.
The expansion of the integral $Y'(t_{m})$ in terms of multiple polylogarithms is
similar to the evaluation of $I'(t_{m})$ in  Eq.~(\ref{iprimes}).
 The result of the calculation is
\ba
\label{B26Cnew}
 Y'(t_m) =Li_{1,1,1,1}\left( \gamma_{3},
-\frac{\gamma_{1}}{\gamma_{3}},-\frac{\gamma_{2}}{\gamma_{1}}, 
\frac{t_{m}}{\gamma_{2}} \right)
+\ln(\gamma_{2}-t_{m}) Li_{1,1,1}\left( \gamma_{3},
-\frac{\gamma_{1}}{\gamma_{3}},-\frac{t_{m}}{\gamma_{1}} \right)\, .
\ea
For the calculation of $Y''(t_m)$  one can make use of (\ref{B23C}). Finally using
Eqs.~(\ref{getB27A}) and (\ref{B26Cnew}) one arrives at the result 
\ba
\label{B27A}
 \int \limits_{0}^{1} dt \frac{ \ln(1-t)   \ln(\gamma_1 + t)
\ln(\gamma_2 - t)} {\gamma_3 -t}=   
-Li_{1,1,1,1}\left(\gamma_{3}, -\frac{\gamma_{1}}{\gamma_{3}}, -
\frac{\gamma_{2}}{\gamma_{1}},  \frac{1}{\gamma_{2}} \right)\nn\\
-Li_{1,1,1,1}\left(\gamma_{3}, \frac{\gamma_{2}}{\gamma_{3}}, -
\frac{\gamma_{1}}{\gamma_{2}}, - \frac{1}{\gamma_{1}} \right)
-\ln(\gamma_{1}+1)Li_{1,1,1}\left(\gamma_{3}, \frac{\gamma_{2}}{\gamma_{3}},  
\frac{1}{\gamma_{2}} \right)\\
-\ln(\gamma_{2}-1)Li_{1,1,1}\left(\gamma_{3}, -\frac{\gamma_{1}}{\gamma_{3}}, -
\frac{1}{\gamma_{1}} \right)
-\ln(\gamma_{1}+1)\ln(\gamma_{2}-1)Li_{1,1}\left(\gamma_{3},  \frac{1}{\gamma_{3}}
\right).\nn
\ea
To obtain the formula for the $L$ function with $\alpha_{2}=0$ we must  only change
$\gamma_{1}, \gamma_{2}$, and $\gamma_{3}$ to
$\alpha_{1}-1\, , \alpha_{3}+1$, and $\alpha_{4}+1$ according to
Eq.~(\ref{changeB27A}):
\ba
\label{fromB27}
L_{-++}(\alpha_{1}, 0, \alpha_{3}, \alpha_{4})=\int \limits_0^1 dy \frac{\ln
(\alpha_1- y) \ln y  \ln(\alpha_{3}+y)} {\alpha_4 + y}=\nn\\
-Li_{1,1,1,1}\left(1+\alpha_{4}, \frac{1-\alpha_{1}}{1+\alpha_{4}},
\frac{1+\alpha_{3}}{1-\alpha_{1}},  \frac{1}{1+\alpha_{3}} \right)
-Li_{1,1,1,1}\left(1+\alpha_{4}   ,
\frac{1+\alpha_{3}}{1+\alpha_{4}},\frac{1-\alpha_{1}}{1+\alpha_{3}},
\frac{1}{1-\alpha_{1}}  \right)\nn\\
-\ln\alpha_{1}Li_{1,1,1}\left(1+\alpha_{4}, \frac{1+\alpha_{3}}{1+\alpha_{4}},
\frac{1}{1+\alpha_{3}}    \right)
-\ln\alpha_{3} Li_{1,1,1}\left(1+\alpha_{4}, \frac{1-\alpha_{1}}{1+\alpha_{4}},
\frac{1}{1-\alpha_{1}}    \right)\nn\\
-\ln\alpha_{1}\ln\alpha_{3} Li_{1,1}\left(1+\alpha_{4}, \frac{1}{1+\alpha_{4}}
\right)\,.\,\,\,\,\,\,\,\,
\ea
For the case $\alpha_{4}=-1$ we calculate the  limit of the rhs of
(\ref{fromB27}) for $\alpha_4 \ra -1$ and obtain
\ba
L_{-++}(\alpha_{1}, 0, \alpha_{3}, -1)=
Li_{2,1,1}\left(1-\alpha_{1} , \frac{1+\alpha_{3}}{1-\alpha_{1}},
\frac{1}{1+\alpha_{3}} \right)\nn\\
+Li_{2,1,1}\left(1+\alpha_{3},\frac{1-\alpha_{1}}{1+\alpha_{3}},
\frac{1}{1-\alpha_{1}}  \right)
+\ln\alpha_{1}Li_{2,1}\left(1+\alpha_{3},  \frac{1}{1+\alpha_{3}}    \right)\nn\\
+\ln\alpha_{3} Li_{2,1}\left(1-\alpha_{1},  \frac{1}{1-\alpha_{1}}    \right)
+\ln\alpha_{1}\ln\alpha_{3}\,  \zeta(2) \,.
\ea

\vglue 1.0cm
              \begin{bf}
\noindent
IV. TRANSFORMATION OF $L_{+++}$ TO MULTIPLE POLYLOGARITHMS
               \end{bf}
\vglue .3cm

In this section we will show that all our $L_{+++}$ functions can be
expressed in terms of multiple polylogarithms. 

\vglue 1.0cm
              \begin{bf}
\noindent
A. General case for the $L_{+++}$ function
               \end{bf}
\vglue .3cm

We now proceed with the transformation of the triple index function $L_{+++}$,
\be \label{getLpppGen}
L_{+++}(\alpha_1,\alpha_2,\alpha_3,\alpha_4)= \int \limits_0^1 dy
\frac{\ln (\alpha_1+y)\ln (\alpha_2+ y)\ln (\alpha_3 + y)}
{\alpha_4 + y}\,  .
\ee
After  changing  the
integration variable $y=-\alpha_{1} t$ we obtain
 \ba
\label{getLpppGeneral}
-\int \limits_{0}^{-1/\alpha_1} dt
\frac{\ln(\alpha_1-\alpha_{1} t) \ln(\alpha_{2}-\alpha_{1}
t)\ln(\alpha_3-\alpha_{1} t)}{\frac{\alpha_{4}}{\alpha_{1}}-t} =
-\int \limits_{0}^{-1/\alpha_1} dt \frac{\ln\alpha_1
\ln(\alpha_{2}-\alpha_{1} t)\ln(\alpha_3+\alpha_{1} t)}
{\frac{\alpha_{4}}{\alpha_{1}}-t}\nn\\
-\int \limits_{0}^{-1/\alpha_1} dt \frac{\ln(1-t) [
\ln\alpha_1+\ln(\frac{\alpha_2}{\alpha_1}-t)] [
\ln\alpha_1+\ln(\frac{\alpha_3}{\alpha_1}- t) ]  }
{\frac{\alpha_4}{\alpha_1}-t}=\nn\\
\ln\alpha_{1} \int \limits_0^1 dy \frac{ \ln (\alpha_2+ y)\ln
(\alpha_3 + y)} {\alpha_4 + y}-\ln^{2}\alpha_1 \int
\limits_{0}^{-1/\alpha_1} dt
\frac{\ln(1-t)}{\frac{\alpha_4}{\alpha_1}-t}\nn\\
-\ln\alpha_1 \int \limits_{0}^{-1/\alpha_1} dt
\frac{\ln(1-t)\ln(\frac{\alpha_2}{\alpha_1}-
t)}{\frac{\alpha_4}{\alpha_1}-t}
-\ln\alpha_1 \int
\limits_{0}^{-1/\alpha_1} dt
\frac{\ln(1-t)\ln(\frac{\alpha_3}{\alpha_1}- t)}{\frac{\alpha_4}{\alpha_1}-t}\nn\\
-\int \limits_{0}^{-1/\alpha_1} dt
\frac{\ln(1-t)\ln(\frac{\alpha_2}{\alpha_1}- t)  
\ln(\frac{\alpha_3}{\alpha_1}- t)
}{\frac{\alpha_4}{\alpha_1}-t}\,.
\hspace{2cm}
 \ea
The first integral on the rhs of (\ref{getLpppGeneral}) has been calculated in
Eq.~(\ref{B23A}). For the
second integral one makes use of the formula (\ref{B28A}) (the
only change is  $\alpha_2\rightarrow \alpha_1$). For the 
evaluation of the third and fourth integrals one uses Eq.
(\ref{B23C}). We are left with the most complicated fifth
integral. Let us consider an integral of the type
\begin{eqnarray}
\int \limits_{0}^{t_m} dt \frac{\ln(1-t)\ln(\gamma_1- t)
\ln(\gamma_2- t) }{\gamma_3-t}\,.
\end{eqnarray}
This integral is an analogue of the integral in Eq.~(\ref{B27}). The calculation
proceeds in a similar way. One obtains the result
\ba
\int \limits_{0}^{t_m} dt \frac{\ln(1-t)\ln(\gamma_1- t)
\ln(\gamma_2- t) }{\gamma_3-t}=-\ln(\gamma_1 -t_m) \ln(\gamma_2 -t_m)
Li_{1,1}\left( \gamma_3, \frac{t_m} {\gamma_3} \right)  \nn
\\
 -\ln(\gamma_2 -t_m) Li_{1,1,1}\left( \gamma_3, \frac{\gamma_1} {\gamma_3},
\frac{t_m}{\gamma_1} \right)
- \ln(\gamma_1-  t_m)
Li_{1,1,1}\left( \gamma_3, \frac{\gamma_2} {\gamma_3}, \frac{t_m}
{\gamma_2} \right)                              \\
- Li_{1,1,1,1}\left( \gamma_3, \frac{\gamma_2} {\gamma_3}, \frac{\gamma_1}
{\gamma_2}, \frac{t_m}{\gamma_1} \right)
- Li_{1,1,1,1} \left( \gamma_3, \frac{\gamma_1} {\gamma_3}, \frac{\gamma_2}
{\gamma_1}, \frac{t_m}{\gamma_2} \right)\, . \nn
\label{B27F}
\ea
Taking into account everything mentioned above for Eq.~(\ref{getLpppGeneral}) we
arrive at the final result for the $L_{+++}$ function,
\ba
\label{LpppGeneral}
L_{+++}(\alpha_1,\alpha_2,\alpha_3,\alpha_4)=
Li_{1,1,1,1}\left(\frac{\alpha_4}{\alpha_1},  \frac{\alpha_2}{\alpha_4},
\frac{\alpha_3}{\alpha_2},-  \frac{1}{\alpha_3}\right)\nn\\
+Li_{1,1,1,1}\left(\frac{\alpha_4}{\alpha_1},  \frac{\alpha_3}{\alpha_4},
\frac{\alpha_2}{\alpha_3}, - \frac{1}{\alpha_2}\right)
+\ln\alpha_{1} Li_{1,1,1}\left(\frac{\alpha_4}{\alpha_2},
\frac{\alpha_3}{\alpha_4}, - \frac{1}{\alpha_3}\right)\nn\\
+\ln(1+\alpha_{2}) Li_{1,1,1}\left(\frac{\alpha_4}{\alpha_1},
\frac{\alpha_3}{\alpha_4}, - \frac{1}{\alpha_3}\right)
+\ln(1+\alpha_{3}) Li_{1,1,1}\left(\frac{\alpha_4}{\alpha_1},
\frac{\alpha_2}{\alpha_4}, - \frac{1}{\alpha_2}\right)\nn\\
+\ln\alpha_{1}\ln\alpha_{2} Li_{1,1}\left(\frac{\alpha_4}{\alpha_3}, -
\frac{1}{\alpha_4}\right)
+\ln\alpha_{1}\ln(1+\alpha_{3}) Li_{1,1}\left(\frac{\alpha_4}{\alpha_2}, -
\frac{1}{\alpha_4}\right) \\
+\ln(1+\alpha_{2})\ln(1+\alpha_{3}) Li_{1,1}\left(\frac{\alpha_4}{\alpha_1}, -
\frac{1}{\alpha_4}\right)
+\ln\alpha_{1}\ln\alpha_{2}\ln\alpha_{3}\ln\left(\frac{\alpha_{4}+1}
{\alpha_{4}}\right) \, .\nn
\ea
For this equation  the conditions \eqr{domain} are assumed. We emphasize that the
arguments of $L_{+++}$ functions occuring in the actual calculation of the massive 
scalar one--loop integrals are not of the most general type as assumed in
the derivation of (\ref{LpppGeneral}). We have nevertheless 
included a discussion of the general
case because Eq.~(\ref{LpppGeneral}) may be useful in other applications.
In the results for the massive scalar integrals one has only the special 
cases where
$\alpha_{1}=\alpha_{2}$ or $\alpha_{1}=\alpha_{3}$
as well as the cases $\alpha_{1}=0$ and/or $\alpha_{2}=0$ and/or $
\alpha_{3}=0$ and/or $\alpha_{4}=\{-1,0\}$.
If some $\alpha$'s coincide with each other Eq.~(\ref{LpppGeneral}) becomes 
simpler. In this case one can also make use of symmetry properties to obtain
simpler relations between the $L_{+++}$ functions and multiple polylogarithms. 
For the cases $\alpha_{1}=0$ and/or   $ \alpha_{2}=0$ and/or   $ 
\alpha_{3}=0$
and/or $\alpha_{4}=\{-1,0\}$ the general formula  (\ref{LpppGeneral}) is no longer
valid and these cases must be studied separately.

\vglue 1.0cm  
              \begin{bf}
\noindent
B. Special cases for the $L_{+++}$ function
               \end{bf}
\vglue .3cm

In the Laurent series expansion of the massive scalar one-loop integrals the
following special cases for the $\alpha_{i}$ are present.

\vglue 1.0cm
              \begin{bf}
\noindent
{\em 1. ${\bf \alpha_{1}=\alpha_{2}}$ or ${\bf \alpha_{1}=\alpha_{3}}$ }
               \end{bf} 
\vglue .3cm

As it was stated in Sec.~II the $L_{+++}$ function is symmetric under the
permutations $\alpha_{i} \lra \alpha_{j}$. 
Therefore, it suffices to consider the case  $\alpha_{1}=\alpha_{2}$.

We must evaluate the integral
\ba
L_{+++}(\alpha_1,\alpha_1,\alpha_3,\alpha_4)=\int \limits_0^1 dy
\frac{\ln^{2} (\alpha_1+y)\ln (\alpha_3 + y)}
{\alpha_4 + y}\,  .
\ea
This integral can be expressed in different  ways. First of all one can directly use
Eq.~(\ref{LpppGeneral})
 replacing $\alpha_{2}$ by $\alpha_{1}$ . The second possibility
is to use symmetry properties. One takes into account the rhs  of
Eq.~(\ref{LpppGeneral}) and
notes that the part with multiple polylogarithms of weight four is symmetric under
the exchange $\alpha_{2}\leftrightarrow\alpha_{3}$. It allows one to reduce the
number of the multiple polylogarithms from two to one.
 First we apply   Eq.~(\ref{LpppGeneral}) for the case $\alpha_{2} =\alpha_{3}$
replacing $\alpha_{3}$ by $\alpha_{2}$. Second we change
 $\alpha_{1}\rightarrow\alpha_{3}$ and $\alpha_{2} \rightarrow \alpha_{1}$. After
these transformations one obtains the following result:
\ba
\label{secversion}
L_{+++}(\alpha_1,\alpha_1,\alpha_3,\alpha_4)=\int \limits_0^1 dy
\frac{\ln^{2} (\alpha_1+y)\ln (\alpha_3 + y)}
{\alpha_4 + y}= \nn\\
+2 Li_{1,1,1,1}\left(\frac{\alpha_4}{\alpha_3},  \frac{\alpha_1}{\alpha_4},1,-
\frac{1}{\alpha_1}\right)
+\ln\alpha_3\,  Li_{1,1,1}\left(\frac{\alpha_4}{\alpha_1},
\frac{\alpha_1}{\alpha_4}, - \frac{1}{\alpha_1}\right)\nn\\
+2 \ln(1+\alpha_1) Li_{1,1,1}\left(\frac{\alpha_4}{\alpha_3},
\frac{\alpha_1}{\alpha_4}, - \frac{1}{\alpha_1}\right)
+\ln\alpha_3 \left[\ln(\alpha_1+1)+\ln\alpha_1  \right]
Li_{1,1}\left(\frac{\alpha_4}{\alpha_1}, - \frac{1}{\alpha_4}\right) \\
+\ln^{2}(1+\alpha_1) Li_{1,1}\left(\frac{\alpha_4}{\alpha_3}, -
\frac{1}{\alpha_4}\right)
+\ln^2\alpha_1 \ln\alpha_3 \ln\left(\frac{\alpha_{4}+1}{\alpha_{4}}\right)\, .\nn
\ea
There is also the third possibility to express
$L_{+++}(\alpha_1,\alpha_1,\alpha_3,\alpha_4) $ in terms of multiple polylogarithms:
\ba
\label{getthirdversion}
\int \limits_0^1 dy
\frac{\ln^{2} (\alpha_1+y)\ln (\alpha_3 + y)}
{\alpha_4 + y}\,  \stackrel{y\rightarrow-\alpha_1 t }{\, =\, } \,
-\int \limits_{0}^{-1/\alpha_1} dt
\frac{\ln^{2}(\alpha_1-\alpha_{1} t) \ln(\alpha_3-\alpha_{1}
t)}{\frac{\alpha_{4}}{\alpha_{1}}-t}=\nn\\
-\int \limits_{0}^{-1/\alpha_1} dt \frac{\left[\ln^2\alpha_1+2\ln\alpha_{1}\ln(1-t)
+\ln^{2}(1-t)\right]\left[ \ln\alpha_{1}+ \ln(\frac{\alpha_3}{\alpha_{1}}-t)\right]}
{\frac{\alpha_{4}}{\alpha_{1}}-t}=\nn\\
+\ln^{3} \alpha_1 \int \limits_0^1
\frac{dy}
{\alpha_4 + y}
+\ln^{2} \alpha_1 \int \limits_0^1 dy
\frac{\ln (\alpha_3 + y)}
{\alpha_4 + y}
-2 \ln^{2}\alpha_{1}\int \limits_{0}^{-1/\alpha_1} dt
\frac{\ln(1-t)}{\frac{\alpha_{4}}{\alpha_{1}}-t}\nn\\
- \ln\alpha_{1}\int \limits_{0}^{-1/\alpha_1} dt
\frac{\ln^{2}(1-t)}{\frac{\alpha_{4}}{\alpha_{1}}-t}
-2\ln\alpha_{1}\int \limits_{0}^{-1/\alpha_1} dt
\frac{\ln(1-t)\ln(\frac{\alpha_3}{\alpha_{1}}-t)}
{\frac{\alpha_{4}}{\alpha_{1}}-t}\nn\\
-\int \limits_{0}^{-1/\alpha_1} dt \frac{\ln^{2}(1-t)\ln(\frac{\alpha_3}{\alpha_{1}}-t)}
{\frac{\alpha_{4}}{\alpha_{1}}-t}\, .
\ea
The first term can be integrated immediately. For the second and third term one uses
Eq.~(\ref{B28}) and Eq.~(\ref{B28A}), respectively.
 The integral of the fourth term can be rewritten as
\ba
 \int \limits_{0}^{-1/\alpha_1} dt
\frac{\ln^{2}(1-t)}{\frac{\alpha_4}{\alpha_1}-t}= 2\int
\limits_{0}^{-1/\alpha_1}\frac{dt_{1}}{\frac{\alpha_4}{\alpha_1}-t_{1}}
\int \limits_{0}^{t_1}\frac{dt_{2}}{1-t_{2}}
\int
\limits_{0}^{t_2}\frac{dt_{3}}{1-t_{3}}=
2Li_{1,1,1}\left(1,\frac{\alpha_{4}}{\alpha_{1}},\frac{-1}{\alpha_{4}} \right)\, .
\ea
The fifth term is  calculable with Eq.~(\ref{B23}). To integrate the last term one
first evaluates the following integral:
\ba
\label{B27E}
\int \limits_{0}^{t_{m}} dt \frac{\ln^{2}(1-t)\ln(\beta_{1}-t)}
{\beta_{2}-t}=-\int \limits_{0}^{t_{m}} dt_{1} \frac{\ln^{2}(1-t_{1})}
{\beta_{2}-t_{1}}
\left \{  \int \limits_{t_{m}}^{t_{1}} +   \int \limits_{\beta_{1}-1}^{t_{m}} \right
\} \frac{dt_{2}}{\beta_1-t_{2}}=\nn\\
\int \limits_{0}^{t_{m}}\frac{dt_{2}}{\beta_1-t_{2}}\int \limits_{0}^{t_2}
dt_{1}\frac{\ln^{2}(1-t_{1})}
{\beta_{2}-t_{1}} +\ln(\beta_{1}-t_{m})\int \limits_{0}^{t_{m}} dt \frac{\ln^{2}(1-t)}
{\beta_{2}-t}=\\
2 Li_{1,1,1,1}\left(1,\beta_{2},\frac{\beta_{1}}{\beta_{2}},\frac{t_{m}}{\beta_{1}} 
\right)+2\ln(\beta_{1}-t_{m})Li_{1,1,1}\left(1,\beta_{2},\frac{t_{m}}{\beta_{2}}\right)\, 
.\nn
\ea
Then to calculate  the last term of Eq.~(\ref{getthirdversion})
one only has to change $\beta_{1}, \beta_{2}$, and $t_{m}$ by the corresponding
combinations of $\alpha_{i}$.
Finally we arrive at the result for the
$L_{+++}(\alpha_1,\alpha_1,\alpha_3,\alpha_4) $ function,
\ba
\label{thirdversion}
L_{+++}(\alpha_1,\alpha_1,\alpha_3,\alpha_4)=
-2Li_{1,1,1,1}\left(1,\frac{\alpha_{4}}{\alpha_{1}},\frac{\alpha_{3}}{\alpha_{4}}, -
\frac{1}{\alpha_{3}}\right)
-2\ln(\alpha_{3}+1) Li_{1,1,1}\left(1,\frac{\alpha_{4}}{\alpha_{1}}, -
\frac{1}{\alpha_{4}}\right)\nn\\
+2\ln \alpha_{1}
Li_{1,1,1}\left(\frac{\alpha_{4}}{\alpha_{1}},\frac{\alpha_{3}}{\alpha_{4}}  , -
\frac{1}{\alpha_{3}}\right)
+2\ln\alpha_{1}\ln(\alpha_{3}+1) Li_{1,1}\left(\frac{\alpha_{4}}{\alpha_{1}}, -
\frac{1}{\alpha_{4}}\right)\nn \\
+\ln^{2}\alpha_{1} Li_{1,1}\left(\frac{\alpha_{4}}{\alpha_{3}}, -
\frac{1}{\alpha_{4}}\right)
+ \ln^{2}\alpha_1\ln\alpha_{3}\ln\left(\frac{\alpha_{4}+1}{\alpha_{4}}   \right)\,
.~~~~~
\ea
This is the third possibility to express
$L_{+++}(\alpha_1,\alpha_1,\alpha_3,\alpha_4) $ function in terms of multiple
polylogaritms.
Each of the Eqs.~(\ref{secversion}) and (\ref{thirdversion}) contains only one
multiple polylogarithm
of weight four and they are both equally acceptable from this point of view. 
One has a free choice to apply any of these equations  for the
required $L$ functions.
The situation with the $L_{+++}(\alpha_1,\alpha_1,\alpha_3,\alpha_4) $ function is
an example of  the statement  that the expansion of the $L$ functions in terms
of multiple polylogarithms is not unique.

\vglue 1.0cm
              \begin{bf}
\noindent
{\em 2. ${\bf \alpha_{1}=0}$ (or ${\bf \alpha_{2}=0}$ or ${\bf \alpha_{3}=0}$)
}
               \end{bf}
\vglue .3cm

For this integral we change the integration variable $y \rightarrow 1-t $,
\ba
L_{+++}(0,\alpha_2,\alpha_3,\alpha_4)= \int \limits_0^1 dy
\frac{\ln y\ln (\alpha_2+ y)\ln (\alpha_3 + y)}
{\alpha_4 + y}=\nn\\
 \int \limits_0^1 dt
\frac{\ln(1-t) \ln (\alpha_2+ 1-t)\ln (\alpha_3 + 1-t)}{\alpha_4 + 1-t}
\ea
and using Eq.~(\ref{B27F}) we arrive at the result
\ba
\label{fromB27F}
L_{+++}(0,\alpha_2,\alpha_3,\alpha_4)=
-Li_{1,1,1,1}\left(1+\alpha_{4}, \frac{1+\alpha_{2}}{1+\alpha_{4}},
\frac{1+\alpha_{3}}{1+\alpha_{2}},  \frac{1}{1+\alpha_{3}} \right)\nn\\
-Li_{1,1,1,1}\left(1+\alpha_{4},
\frac{1+\alpha_{3}}{1+\alpha_{4}},\frac{1+\alpha_{2}}{1+\alpha_{3}},
\frac{1}{1+\alpha_{2}}  \right)
-\ln\alpha_{2}Li_{1,1,1}\left(1+\alpha_{4}, \frac{1+\alpha_{3}}{1+\alpha_{4}},
\frac{1}{1+\alpha_{3}}    \right)\nn\\
-\ln\alpha_{3} Li_{1,1,1}\left(1+\alpha_{4}, \frac{1+\alpha_{2}}{1+\alpha_{4}},
\frac{1}{1+\alpha_{2}}    \right)
-\ln\alpha_{2}\ln\alpha_{3} Li_{1,1}\left(1+\alpha_{4}, \frac{1}{1+\alpha_{4}}
\right).\,\,
\ea

\vglue 1.0cm
              \begin{bf}
\noindent
{\em 3. ${\bf \alpha_{1}=\alpha_{2}=0}$ }
               \end{bf}
\vglue .3cm

To calculate this integral we again change the integration variable $y 
\rightarrow 1-t $,
\ba
L_{+++}(0,0,\alpha_3,\alpha_4)= \int \limits_0^1 dy
\frac{\ln^{2} y \ln (\alpha_3 + y)}
{\alpha_4 + y}=\nn\\
 \int \limits_0^1 dt
\frac{\ln^{2}(1-t) \ln (\alpha_3 + 1-t)}{\alpha_4 + 1-t} .
\ea
For the last integral we use Eq.~(\ref{B27E}). An additional simplification can be
done if one notes that
\ba
Li_{1,1,1}\left(1,\alpha_{4}+1,\frac{1}{\alpha_{4}+1}\right)=-{\rm 
Li}_{3}\left(-\frac{1}{\alpha_4}  \right) .
\ea
Finally one has
\ba
\label{fromB27E}
L_{+++}(0,0,\alpha_3,\alpha_4)=
2Li_{1,1,1,1}\left(1,\alpha_{4}+1,\frac{\alpha_{3}+1}
{\alpha_{4}+1},\frac{1}{\alpha_{3}+1} \right)
-2\ln\alpha_{3}{\rm Li}_3 \left(-\frac{1}{\alpha_{4}}\right).
\ea

\vglue 1.0cm
              \begin{bf}
\noindent
{\em 4. ${\bf \alpha_{1}=\alpha_{2}=0, \z  \alpha_4=-1 }$  (or
   ${\bf \alpha_2=\alpha_{3}=0, \z  \alpha_{4}=-1 }$)  }
               \end{bf}
\vglue .3cm

In this case one should calculate the limit of the rhs of (\ref{fromB27E}) for
$t_{m}=1$ and $  \alpha_{4}\rightarrow -1$.
After this procedure one obtains
\ba
L_{+++}(0,0,\alpha_3,-1)=-2 Li_{1,2,1}\left(1,\alpha_{3}+1,\frac{1}{\alpha_{3}+1} \right)
-2\ln\alpha_{3}\zeta(3).
\ea
For the case $\alpha_2=\alpha_{3}=0$ and $\alpha_{4}=-1$ one can use the same formula.
The only change is $\alpha_{3}\rightarrow\alpha_{1}$.

\vglue 1.0cm
              \begin{bf}
\noindent
{\em 5. ${\bf \alpha_{1}=0, \z \alpha_{4}=-1}$ }
               \end{bf}
\vglue .3cm

To obtain the solution for these values of the $\alpha_{i}$ we must find the
limit  of the rhs of (\ref{fromB27F}) for $\alpha_{4}\rightarrow -1$.
After taking the limit one arrives at the result
\ba
L_{+++}(0,\alpha_2,\alpha_3,-1)=
+Li_{2,1,1}\left(1+\alpha_{2}, \frac{1+\alpha_{3}}{1+\alpha_{2}},
\frac{1}{1+\alpha_{3}} \right)\nn\\
+Li_{2,1,1}\left(1+\alpha_{3}, \frac{1+\alpha_{2}}{1+\alpha_{3}},
\frac{1}{1+\alpha_{2}}  \right)
+\ln\alpha_{2}Li_{2,1}\left(1+\alpha_{3},  \frac{1}{1+\alpha_{3}}    \right)\nn\\
+\ln\alpha_{3} Li_{2,1}\left(1+\alpha_{2},   \frac{1}{1+\alpha_{2}}    \right)
+\ln\alpha_{2}\ln\alpha_{3} \zeta(2) .
\ea

\vglue 1.0cm
              \begin{bf}
\noindent
V. TRANSFORMATION OF $L_{+}$ TO MULTIPLE POLYLOGARITHMS 
               \end{bf}  
\vglue .3cm

In this section we will show that all our $L_{+}$ functions can be
expressed in terms of multiple polylogarithms.

\vglue 1.0cm
              \begin{bf}
\noindent
A. General case for the $L_{+}$ function
               \end{bf}
\vglue .3cm

Here we derive the general formula for  the single index $L_{+}$ function
Eq.~(\ref{Lpfunction}),
\be
L_{+}(\alpha_1,\alpha_2,\alpha_3,\alpha_4)=
\int\limits_0^1 dy\, \frac{\ln (\alpha_1+y)} {\alpha_4+y} {\rm
Li}_2(\alpha_2+\alpha_3 y).
\ee
After changing the integration variable $y\rightarrow(t-\alpha_2)/\alpha_3$ one gets
\be
L_{+}=\int \limits_{\alpha_2}^{\alpha_2+\alpha_3} \frac{dt}{\alpha_3}
\frac{\ln (\alpha_1+ \frac{t-\alpha_2}{\alpha_3})}
{\alpha_4+\frac{t-\alpha_2}{\alpha_3}} {\rm Li}_2(t) =
\int \limits_{\alpha_2}^{\alpha_2+\alpha_3}
dt \frac{-\ln\alpha_3 + \ln (\alpha_1
\alpha_3 - \alpha_2 + t)}{\alpha_3 \alpha_4 - \alpha_2 + t} {\rm
Li}_2(t).
\ee
The integration interval can be split into two pieces, $[\alpha_2, 0]$ and
$[0, \alpha_2 + \alpha_3]$. One can then write $L_{+}$ as a sum of four
terms,
\be
\label{sum4}
L_{+} =- \ln \alpha_3 \left\{ \int
\limits_{0}^{\alpha_2+\alpha_3}  - \int \limits_{0}^{\alpha_2} \,\, \right\}
\frac{dt} {\gamma+t} {\rm Li}_{2}(t)
+ \left\{ \int \limits_{0}^{\alpha_2+\alpha_3}
                              - \int \limits_{0}^{\alpha_2} \,\,\right\}
dt \frac{\ln(\alpha+t)} {\gamma+t} {\rm Li}_{2}(t),
\ee
where we have introduced the notation
\be
\label{algama}
\alpha = \alpha_1 \alpha_3 - \alpha_2, {\rm  \hspace{0.4in}}
\gamma = \alpha_3 \alpha_4 - \alpha_2.
\ee
Looking at  Eq.~(\ref{sum4}) it is clear that there are only two different
types of integrals to be dealt with,
\be
\label{types}
\int \limits_{0}^{t_m} \frac{dt} {\gamma+t} {\rm Li}_{2}(t)
{\rm \hspace{0.4in}  and}    {\rm \hspace{0.4in}}
\int \limits_{0}^{t_m} dt \frac{\ln(\alpha+t)} {\gamma+t} {\rm Li}_{2}(t).
\ee   
The upper limits are $t_m = \alpha_2 + \alpha_3$ or $t_m = \alpha_2$.
The first integral can be evaluated analytically in terms
of standard logarithms and classical polylogarithms up to ${\rm Li}_3$. However,
the same integral can also be expressed in terms of multiple polylogarithms via
the integral representation (\ref{intrepr}), e.g.,
\be
\label{multi1}
\int \limits_{0}^{t_m}\frac{dt}{\gamma+t}{\rm Li}_{2}(t)=
\int \limits_{0}^{t_m}\frac{dt_{1}}{\gamma+t_{1}}  
\int \limits_{0}^{t_1} \frac{dt_{2}}{t_2}
\int \limits_{0}^{t_2} \frac{dt_3}{1-t_3} =
-Li_{2,1}\left(-\gamma,\frac{-t_m}{\gamma}\right).
\ee
We now deal with
the second integral in (\ref{types}). Consider the following multiple
polylogarithm of weight four:
\ba
Li_{2,1,1} \left( -\gamma, \frac{\alpha}{\gamma},\frac{t_m}{-\alpha} \right)=
\int \limits_{0}^{t_m}  \frac{dt_{2}}{-\alpha-t_{2}}  \int \limits_{0}^{t_{2}}
\frac{dt_{1}}{-\gamma-t_{1}}{\rm Li}_{2}(t_{1})   =
\int \limits_{0}^{t_m}\frac{dt_{1}}{\gamma+t_{1}}{\rm Li}_{2}(t_{1}) \int
\limits_{t_{1}}^{t_m}  \frac{dt_{2}}{\alpha+t_{2}}=    \nn \\
\int \limits_{0}^{t_m}\frac{dt_{1}}{\gamma+t_{1}}{\rm Li}_{2}(t_{1})
\ln(\alpha+t_m) - \int
\limits_{0}^{t_m}\frac{dt_{1}}{\gamma+t_{1}}{\rm Li}_{2}(t_{1}) \ln(\alpha+t_{1}).~~~
\ea
In the first step we have used the usual trick to change the order of
integration.
As already noted before [see Eq.~(\ref{multi1})] the first term on the second line
can be expressed through a multiple polylogarithm of weight three. Thus one has
\be
\label{multi2}
 \int \limits_{0}^{t_m} dt \frac{\ln(\alpha+t)} {\gamma+t} {\rm Li}_{2}(t)
=
 - Li_{2,1,1}\left(-\gamma, \frac{\alpha}{\gamma},\frac{t_m}{-\alpha}\right)  -
Li_{2,1}\left(-\gamma,\frac{-t_m}{\gamma}\right) \ln (\alpha+t_m).
\ee
Finally, substituting Eqs.~(\ref{multi1}) and (\ref{multi2}) into
Eq.~(\ref{sum4}) we arrive at the desired relation
\ba
\label{LpGeneral}
L_{+}(\alpha_1,\alpha_2,\alpha_3,\alpha_4)=
Li_{2,1,1}\left(\alpha_{2}-\alpha_{3}\alpha_{4},
\frac{\alpha_{2}-\alpha_{1}\alpha_{3} }{\alpha_{2}-\alpha_{3}\alpha_{4}},
\frac{\alpha_2}{\alpha_{2} -\alpha_{1}\alpha_{3}  }\right) \nn\\
-Li_{2,1,1}\left(\alpha_{2}-\alpha_{3}\alpha_{4},
\frac{\alpha_{2}-\alpha_{1}\alpha_{3} }{\alpha_{2}-\alpha_{3}\alpha_{4}},
\frac{\alpha_2+\alpha_{3}}{\alpha_{2} -\alpha_{1}\alpha_{3}  }\right)
+\ln\alpha_{1} Li_{2,1} \left( \alpha_{2}-\alpha_{3}\alpha_{4}  ,
\frac{\alpha_2}{\alpha_{2} -\alpha_{3}\alpha_4}  \right) \\
-\ln(\alpha_{1}+1)
Li_{2,1} \left( \alpha_{2}-\alpha_{3}\alpha_{4}  ,
\frac{\alpha_2+\alpha_{3}}{\alpha_{2} -\alpha_{3}\alpha_{4}}  \right)  .\nn
\ea
We should note that, similar to Eq.~(\ref{B27General}), the conditions~(\ref{domain})
are assumed  for the variables $\alpha_{i}$.
Also, one cannot directly use Eq.~(\ref{LpGeneral}) if
$\alpha_{2}-\alpha_{3}\alpha_{4}=0$ or $\alpha_{2}-\alpha_{1}\alpha_{3}=0$.
However,  in the results for the massive scalar integrals precisely these special cases  
appear, as well as the cases
where $\alpha_{1}=0$ and/or $\alpha_{2}=0$ and/or $ \alpha_{3}=0$ and/or
$\alpha_{4}=\{-1,0\}$.
In such cases the general formula~(\ref{LpGeneral}) is no longer valid and these
cases must be studied separately.


\vglue 1.0cm
              \begin{bf}
\noindent
B. Special cases for the $L_{+}$ function
               \end{bf}
\vglue .3cm

In the Laurent series expansion of the massive scalar one-loop integrals the
following special cases appear
for the  arguments of the $L_{+}$ functions:

\vglue 1.0cm
              \begin{bf}
\noindent
{\em 1. ${\bf \alpha_{2}-\alpha_{3}\alpha_{4}=0 }$
                 (or ${\bf \alpha_{2}-\alpha_{1}\alpha_{3}=0 }$) }
               \end{bf}
\vglue .3cm

In this case one must find the limit of the rhs of Eq.~(\ref{LpGeneral}) for
$\alpha_{2}\rightarrow \alpha_{3}\alpha_{4} $.
First we rewrite the rhs of Eq.~(\ref{LpGeneral}) in terms of multidimensional
integrals via the definition~(\ref{intrepr}). Second we replace
$\alpha_{2}$ by $\alpha_{3}\alpha_{4}$. We finally again use the
definition~(\ref{intrepr}) to obtain the result
\ba
L_{+}(\alpha_1, \alpha_3 \alpha_4  ,\alpha_3,\alpha_4)=
-Li_{3,1}\left(\alpha_{3}(\alpha_{4}-\alpha_{1}),\frac{\alpha_4}{\alpha_{4}
-\alpha_{1}  }\right)\nn \\
+ Li_{3,1}\left(\alpha_{3}(\alpha_{4}-\alpha_{1}),\frac{\alpha_4+1}{\alpha_{4}
-\alpha_{1}  }\right)
-\ln\alpha_{1}{\rm Li}_{3}(\alpha_{3}\alpha_{4})+\ln(\alpha_{1}+1){\rm
Li_3}\left(\alpha_3(\alpha_{4}+1)\right) .
\ea

When $\alpha_{2}-\alpha_{1}\alpha_{3}=0$ one must find the limit of the rhs 
of Eq.~(\ref{LpGeneral}) for
$\alpha_{2}\rightarrow \alpha_{1}\alpha_{3} $. We again rewrite the rhs of
Eq.~(\ref{LpGeneral}) 
in terms of multidimensional 
integrals. We then replace $\alpha_{2}$ by $\alpha_{1}\alpha_{3}$ and use the 
definition~(\ref{intrepr}). We arrive at the result
\ba
L_{+}(\alpha_1, \alpha_1 \alpha_3  ,\alpha_3,\alpha_4)=
 - Li_{2,2}\left(\alpha_{3}(\alpha_{1}-\alpha_{4}),\frac{\alpha_1}{\alpha_{1}
-\alpha_{4}  }\right)\nn \\
+Li_{2,2}\left(\alpha_{3}(\alpha_{1}-\alpha_{4}),\frac{\alpha_1+1}{\alpha_{1}
-\alpha_{4}  }\right)
+\ln\alpha_1 Li_{2,1}\left(\alpha_{3}(\alpha_{1}-\alpha_{4}),\frac{\alpha_1}
{\alpha_{1}-\alpha_{4}  }\right) \\
-\ln(\alpha_{1}+1)
Li_{2,1}\left(\alpha_{3}(\alpha_{1}-\alpha_{4}),\frac{\alpha_1+1}
{\alpha_{1}-\alpha_{4}  }\right) \nn .
\ea


\vglue 1.0cm
              \begin{bf}
\noindent
{\em 2. ${\bf \alpha_1=0 }$  }
               \end{bf}
\vglue .3cm

Unfortunately in this case one cannot use Eq.~(\ref{LpGeneral}) for $\alpha_1=0$
because one is immediately faced with the problem of a logarithmic
infinity. One must find another algorithm to express the $L_{+}(0,
\alpha_2,\alpha_3,\alpha_4)$ function in terms of multiple polylogarithms.
After  changing  the integration variable $y\rightarrow 1-t$ one gets
\ba
\label{togetChangeLp1}
\int\limits_0^1 dy\, \frac{\ln y} {\alpha_4+y}
{\rm Li_2}(\alpha_2+\alpha_3 y)= \int\limits_0^1 dt\, \frac{\ln (1-t)} 
{\alpha_4+1-t} {\rm Li}_2(\alpha_2+\alpha_3 -\alpha_{3}t)=\nn\\   
\int \limits_0^1 dt_{1} \frac{\ln (1-t_{1})} {\alpha_4+1-t_{1}}
\int \limits_{\alpha_2/ \alpha_3+1}^{t_{1}} dt_{2}
\frac{\ln(1-\alpha_{2}-\alpha_{3}+\alpha_{3} t_{2})
}{\frac{\alpha_{2}}{\alpha_{3}}+1-t_{2}}=\nn\\  
\nn\\
\int \limits_0^1 dt_{1} \frac{\ln (1-t_{1})} {\alpha_4+1-t_{1}} \left\{\int
\limits_{1}^{t_{1}}+\int \limits_{\alpha_2/ \alpha_3+1}^{1}    \right\}   
dt_{2} \frac{\ln(1-\alpha_{2}-\alpha_{3}+\alpha_{3} t_{2})
}{\frac{\alpha_{2}}{\alpha_{3}}+1-t_{2}}
=\\
\nn\\
-\int \limits_{0}^{1} dt_{2} \frac{\ln(1-\alpha_{2}-\alpha_{3}+\alpha_{3} t_{2})
}{\frac{\alpha_{2}}{\alpha_{3}}+1-t_{2}}
 \int \limits_0^{t_{2}} dt_{1} \frac{\ln (1-t_{1})} {\alpha_4+1-t_{1}}
-{\rm Li}_{2}(\alpha_{2}) Li_{1,1}\left(\alpha_{4}+1,\frac{1}{\alpha_{4}+1  } 
\right) .\nn
\ea
The last integral is an analogue of $I'(t_{m})$ in Eq.~(\ref{iprimes}). First one
notes that
\ba
\label{smallhelp}
\int \limits_0^{t_{2}} dt_{1} \frac{\ln (1-t_{1})} {\alpha_4+1-t_{1}}=
-Li_{1,1}\left(\alpha_{4}+1,\frac{t_{1}}{\alpha_{4}+1  }\right) .
\ea
Then one considers the following chain of transformations:
\ba
\label{getB26F}
\int \limits_{0}^{1}     \frac{dt_{2}}{1-\alpha_{2}-\alpha_{3}+\alpha_{3} t_{2}}
 \int \limits_{0}^{t_{2}}\frac{dt_{1}}{\frac{\alpha_{2}}{\alpha_{3}}+1-t_{1}}
Li_{1,1}\left(\alpha_{4}+1,\frac{t_{1}}{\alpha_{4}+1  }\right)=\nn\\
\int \limits_{0}^{1} \frac{dt_{1}}{\frac{\alpha_{2}}{\alpha_{3}}+1-t_{1}}  
Li_{1,1}\left(\alpha_{4}+1,\frac{t_{1}}{\alpha_{4}+1  }\right)
\int \limits_{t_{1}}^{1}     \frac{dt_{2}}{1-\alpha_{2}-\alpha_{3}+\alpha_{3} t_{2}}=\nn\\
\frac{1}{\alpha_{3}}\ln(1-\alpha_{2})   
\int \limits_{0}^{1} \frac{dt_{1}}{\frac{\alpha_{2}}{\alpha_{3}}+1-t_{1}}
Li_{1,1}\left(\alpha_{4}+1,\frac{t_{1}}{\alpha_{4}+1  }\right)\\
-\frac{1}{\alpha_{3}}
\int \limits_{0}^{1}dt_{1} \frac{\ln(1-\alpha_{2}-\alpha_{3}+\alpha_{3} t_{1})
}{\frac{\alpha_{2}}{\alpha_{3}}+1-t_{1}}
Li_{1,1}\left(\alpha_{4}+1,\frac{t_{1}}{\alpha_{4}+1  }\right) \nn
\ea
Using Eq.~(\ref{smallhelp}) we see that the  last integral is exactly the integral
required in Eq.~(\ref{togetChangeLp1}).
The initial integral of Eq.~(\ref{getB26F}) and the first integral of the rhs of
Eq.~(\ref{getB26F}) can be expressed in terms of
multiple polylogarithms due to the definition (\ref{intrepr}). Finally for the
$L_{+}(0, \alpha_2,\alpha_3,\alpha_4)$ function we obtain
\ba
\label{changeLp1}
L_{+}(0, \alpha_2,\alpha_3,\alpha_4)=
Li_{1,1,1,1}\left(\alpha_{4}+1,\frac{\alpha_2+\alpha_3}{\alpha_{3}(\alpha_4 +1)  },
\frac{\alpha_2+\alpha_3-1}{ \alpha_2+\alpha_3  }, \frac{\alpha_3}{
\alpha_2+\alpha_3-1  }  \right) ~~~\\
+\ln(1-\alpha_{2})Li_{1,1,1}\left(\alpha_{4}+1,\frac{\alpha_2+\alpha_3}
{\alpha_{3}(\alpha_4 +1)  }, \frac{\alpha_3}{ \alpha_2+\alpha_3  }  \right)
-{\rm Li}_{2}(\alpha_{2})Li_{1,1}\left(\alpha_{4}+1,\frac{1}{\alpha_4 +1  }  \right) . \nn
\ea   

\vglue 1.0cm
              \begin{bf}
\noindent
{\em 3. ${\bf \alpha_1=0,   \z  \alpha_{4}=-1 }$  }
               \end{bf}
\vglue .3cm

For these values  of the $\alpha_{i}$ one uses Eq.~(\ref{changeLp1}) to calculate
the limit of the rhs for $\alpha_{4}\rightarrow -1$.
One arrives at the result
\ba
L_{+}(0, \alpha_2,\alpha_3,-1)=-Li_{2,1,1}\left(\frac{\alpha_2+\alpha_3}{\alpha_{3} },
\frac{\alpha_2+\alpha_3-1}{ \alpha_2+\alpha_3  }, \frac{\alpha_3}{
\alpha_2+\alpha_3-1  }  \right)\\
-\ln(1-\alpha_{2})Li_{2,1}\left(\frac{\alpha_2+\alpha_3}{\alpha_{3}  },
\frac{\alpha_3}{ \alpha_2+\alpha_3  }  \right)
+ {\rm Li}_{2}(\alpha_{2}) \zeta(2) . \nn
\ea

\vglue 1.0cm
              \begin{bf}
\noindent
{\em 4. ${\bf \alpha_1=0,  \z  \alpha_{2}+\alpha_{3}=1}$ (and
                                                    ${\bf \alpha_{4}=-1}$)  }
               \end{bf}
\vglue .3cm

If one takes a look at Eq.~(\ref{changeLp1}) one realizes that there is a problem if
$\alpha_{2}+\alpha_{4}=1$.
To express the $L_{+}$ function for this configuration of the $\alpha_{i}$ the limit
of the rhs of ~(\ref{changeLp1})  
for $\alpha_{2}\rightarrow 1 - \alpha_{3}$ must be found.
The result is
\ba
L_{+}(0,1-\alpha_{3},\alpha_{3} ,\alpha_{4})=
-Li_{1,1,2}\left(\alpha_{4}+1,\frac{1}{\alpha_{3}(\alpha_{4}+1)  }, \alpha_{3}  \right)\\
+\ln\alpha_{3}Li_{1,1,1}\left(\alpha_{4}+1,\frac{1}{\alpha_{3}(\alpha_{4}+1)  },
\alpha_{3}  \right)
-{\rm Li}_{2}(1-\alpha_{3})Li_{1,1}\left(\alpha_{4}+1,\frac{1}{\alpha_{4}+1  }   
\right) . \nn
\ea
For the case $\alpha_1=0$,  $\alpha_{2}+\alpha_{3}=1$, and $\alpha_{4}=-1$ one must
find in addition the limit for  $\alpha_{4}\rightarrow-1$.
One arrives at the result
\ba
L_{+}(0,1-\alpha_{3},\alpha_{3} ,-1)=
Li_{2,2}\left(\frac{1}{\alpha_{3}  }, \alpha_{3}  \right)
-\ln\alpha_{3}Li_{2,1}\left(\frac{1}{\alpha_{3} }, \alpha_{3}  \right)
+\zeta(2) {\rm Li_{2}}(1-\alpha_{3}) .
\ea

\vglue 1.0cm
              \begin{bf}
\noindent
{\em 5. ${\bf \alpha_1=0,  \z   \alpha_{2}=-\alpha_{3} }$  }
               \end{bf}
\vglue .3cm

To obtain the result for this case one must calculate the limit of the rhs
of~(\ref{changeLp1}) for
$\alpha_{3}\rightarrow-\alpha_{2}$.
After taking  the limit one has
\ba
L_{+}(0,\alpha_{2},-\alpha_{2} ,\alpha_{4})= 
-Li_{1,1,2}\left(\frac{\alpha_{2}}{\alpha_{2}-1  },- \alpha_{4},
-\frac{1}{\alpha_{4}}  \right)\nn\\
+\ln(1-\alpha_{2})Li_{1,2}\left(- \alpha_{4}, -\frac{1}{\alpha_{4}}  \right)
+{\rm Li}_{2}(\alpha_{2}) {\rm Li}_{2}\left(-\frac{1}{\alpha_{4}}\right).
\ea

\vglue 1.0cm
              \begin{bf}
\noindent
{\em 6. ${\bf \alpha_1=0, \z  \alpha_{2}=0}$  }
               \end{bf}
\vglue .3cm

For this case one can directly use Eq.~(\ref{changeLp1}),
\ba
\label{LpChange7V2}
L_{+}(0, 0,\alpha_3,\alpha_4)=Li_{1,1,1,1}\left(\alpha_{4}+1,\frac{1}{\alpha_4 +1  },
\frac{\alpha_3-1}{\alpha_3  }, \frac{\alpha_3}{\alpha_3-1  }  \right).
\ea
But there is also another very simple possibility. We first change the integration
variable $y\rightarrow t/\alpha_{3}$,
\ba
\int\limits_0^1 dy\, \frac{\ln y} {\alpha_4+y} {\rm Li_2}(\alpha_3 y)=
\int\limits_0^{\alpha_{3}} dt  \frac{\ln(t / \alpha_{3}) } {\alpha_{3}\alpha_4+t}
{\rm Li_2}(t)=
\int\limits_0^{\alpha_{3}}   \frac{dt_{1} } {\alpha_{3}\alpha_4+t_{1}} {\rm Li_2}(t_{1})
\int\limits_{\alpha_3}^{t_{1}}  \frac{dt_{2} } {t_{2}}=\nn\\
-\int\limits_{0}^{\alpha_{3}}  \frac{dt_{2} } {t_{2}}
\int\limits_0^{t_{2}}   \frac{dt_{1} } {\alpha_{3}\alpha_4+t_{1}} {\rm Li_2}(t_{1}) =
\int\limits_{0}^{\alpha_{3}}  \frac{dt_{2} } {t_{2}}
\int\limits_0^{t_{2}}   \frac{dt_{1} } {-\alpha_{3}\alpha_4+t_{1}}
\int\limits_{0}^{t_{1}}  \frac{dt_{3} } {t_{3}}  \int\limits_{0}^{t_{3}}
\frac{dt_{4} } {1-t_{4}}.
\ea
Now using the definition~(\ref{intrepr}) we obtain the result
\ba
\label{LpChange7V1}   
L_{+}(0,
0,\alpha_3,\alpha_4)=Li_{2,2}\left(-\alpha_{3}\alpha_{4},\frac{-1}{\alpha_4}\right).
\ea
The reader has a free choice to use either formula (\ref{LpChange7V2}) or
(\ref{LpChange7V1}). Both equations contain
multiple polylogarithm of weight four. The depth of the multiple polylogarithm in
Eq.~(\ref{LpChange7V1}) is two against
four in Eq.~(\ref{LpChange7V2}). For $\alpha_{4}=-1$ Eq.~(\ref{LpChange7V1}) can be
directly used. However, in the case of Eq.~(\ref{LpChange7V2})
one must first calculate the limit for $\alpha_{4}\ra -1$.

\vglue 1.0cm
              \begin{bf}
\noindent
{\em 7. ${\bf \alpha_1=0,    \z  \alpha_{2}=1 }$  }
               \end{bf}
\vglue .3cm

Unfortunately, in this case one cannot use Eq.~(\ref{changeLp1}) because of the term
$\ln(1-\alpha_{2})$.
To express this $L_{+}$ function in terms of multiple polylogarithms we first make use of 
a standard relation between dilogs with arguments $x$ and $1-x$  
for the function ${\rm Li}_{2}$ under the sign of the integral:
\ba
\label{getChangeLp8}
\int\limits_0^1 dy \frac{\ln y} {\alpha_4+y} {\rm Li_2}(1+\alpha_3 y)=
\int\limits_0^1 dy \frac{\ln y} {\alpha_4+y} [\zeta(2)
-\ln(-\alpha_{3}y)\ln(1+\alpha_{3}y)-{\rm Li_2}(-\alpha_3 y)]=\nn\\
\zeta(2)\int\limits_0^1 dy\, \frac{\ln y} {\alpha_4+y}
-\int\limits_0^1 dy  \frac{\ln y} {\alpha_4+y}{\rm Li_2}(-\alpha_3 y)
-\int\limits_0^1 dy \frac{\ln y  [ \ln(-\alpha_{3})+\ln y     ] \ln(1+\alpha_{3}y)
} {\alpha_4+y} =\nn\\
\zeta(2){\rm Li}_{2}\left(-\frac{1}{\alpha_{4}}   \right)
-Li_{1,1,1,1}\left(\alpha_{4}+1,\frac{1}{\alpha_4 +1  },
\frac{\alpha_3+1}{\alpha_3  }, \frac{\alpha_3}{\alpha_3+1  }  \right)~~~ \\
-\ln(-\alpha_{3})   \int\limits_0^1 dy\, \frac{\ln y  \ln(1+\alpha_{3}y)   } {\alpha_4+y}
-\int\limits_0^1 dy \frac{\ln^{2} y  \ln(1+\alpha_{3}y)   } {\alpha_4+y}\, ,~~~\nn
\ea
where the $Li_{1,1,1,1}$ function was obtained with the help of 
Eq.~(\ref{LpChange7V2}).
To obtain the last integral in Eq.~(\ref{getChangeLp8}) one proceeds as follows:
\ba
\label{helpLp8a}
\int\limits_0^1 dy \frac{\ln^{2} y  \ln(1+\alpha_{3}y)   } {\alpha_4+y} \,
\stackrel{y\rightarrow1-t }{\, =\, } \,
\int\limits_0^1 dt \frac{\ln^{2} (1-t)  \ln(1+\alpha_{3}-\alpha_{3}t)   }
{\alpha_4+1-t}=\nn\\
\int\limits_0^1 dt_{1} \frac{\ln^{2} (1-t_{1})    } {\alpha_4+1-t_{1}}
\int\limits_1^{t_{1}} \frac{-\alpha_{3} dt_{2}}{1+\alpha_{3}-\alpha_{3}t_{2}   }=
\int\limits_0^{1} \frac{ dt_{2}}{\frac{1}{\alpha_{3}}+1-t_{2}
}\int\limits_0^{t_{2}} dt_{1} \frac{\ln^{2} (1-t_{1})    } {\alpha_4+1-t_{1}}=\nn\\
2 \int\limits_0^{1} \frac{ dt_{2}}{\frac{1}{\alpha_{3}}+1-t_{2}
}\int\limits_0^{t_{2}}\frac{dt_{1}   } {\alpha_4+1-t_{1}}
\int\limits_0^{t_{1}}\frac{dt_{3}}{1-t_{3}}
\int\limits_0^{t_{3}}\frac{dt_{4}}{1-t_{4}}= \\
2 Li_{1,1,1,1}\left(1,\alpha_{4}+1,\frac{\alpha_3+1}{\alpha_3(\alpha_{4}+1 )  },
\frac{\alpha_3}{\alpha_3+1  }  \right). \nn
\ea
Similarly one can evaluate the remaining  integral
\ba
\label{helpLp8b}
\int\limits_0^1 dy \frac{\ln y  \ln(1+\alpha_{3}y)   } {\alpha_4+y}=
-Li_{1,1,1}\left(\alpha_{4}+1,\frac{\alpha_3+1}{\alpha_3(\alpha_{4}+1 )  },
\frac{\alpha_3}{\alpha_3+1  }  \right).
\ea
Now combining Eqs.~(\ref{getChangeLp8}), (\ref{helpLp8a}), and (\ref{helpLp8b}) one
arrives at the result
\ba
\label{ChangeLp8}
L_{+}(0,1,\alpha_{3},\alpha_{4})=
-2 Li_{1,1,1,1}\left(1,\alpha_{4}+1,\frac{\alpha_3+1}{\alpha_3(\alpha_{4}+1 )  },
\frac{\alpha_3}{\alpha_3+1  }  \right)\nn\\
-Li_{1,1,1,1}\left(\alpha_{4}+1,\frac{1}{\alpha_4 +1  },
\frac{\alpha_3+1}{\alpha_3  }, \frac{\alpha_3}{\alpha_3+1  }  \right)\\
+\ln(-\alpha_{3})
Li_{1,1,1}\left(\alpha_{4}+1,\frac{\alpha_3+1}{\alpha_3(\alpha_{4}+1 )  },
\frac{\alpha_3}{\alpha_3+1  }  \right)
+\zeta(2){\rm Li}_{2}\left(-\frac{1}{\alpha_{4}}   \right). \nn
\ea

\vglue 1.0cm
              \begin{bf}
\noindent
{\em 8. ${\bf \alpha_1=0,   \z    \alpha_{2}=-\alpha_{3}=1 }$  }
               \end{bf}
\vglue .3cm

For  these values   of the $\alpha_{i}$ we must find the limit of the rhs of
Eq.~(\ref{ChangeLp8}) for $\alpha_{3}\rightarrow -1$. After taking the limit 
we obtain
\ba
L_{+}(0,1,-1,\alpha_{4})=
Li_{1,1,2}\left(\alpha_{4}+1,\frac{1}{\alpha_{4}+1  }, 1  \right)\nn\\
+2 Li_{1,1,2}\left(1,\alpha_{4}+1,\frac{1}{\alpha_{4}+1  }  \right)
+\zeta(2){\rm Li}_{2}\left(-\frac{1}{\alpha_{4}}   \right).
\ea

\vglue 1.0cm
              \begin{bf}
\noindent
VI. CONCLUSIONS
               \end{bf}
\vglue .3cm

We have presented all the necessary relations to transform the 
$L$ functions [as defined in Eqs.~\eqr{Lfunction} and \eqr{Lpfunction}] that occur in
our ${\cal O}(\varepsilon^2)$ results \cite{KMR} for the Laurent series expansion 
of massive scalar one-loop integrals to multiple polylogarithms.
We have used these relations to transform our results on massive one-loop
integrals involving $L$ functions to corresponding results involving
multiple polylogarithms. 
The multiple polylogarithms results are readily available in electronic form 
\cite{epaps}.

Despite of the fact that the relations between the $L$ functions and the
multiple polylogarithms have been derived having the massive scalar one-loop 
integrals in mind they can also be used in a more general setting. In fact, any 
definite integral given by 
\ba
 \int\limits_A^B  \frac{ \ln (a_{1}+b_{1} x ) \ln (a_{2}+b_{2} x)  \ln
(a_{3}+b_{3}x)
dx } { a_{4}+b_{4}x   }~  {\rm or} ~
 \int\limits_A^B  \frac{ \ln (a_{1}+b_{1} x ) {\rm Li}_{2}(a_{2}+b_{2} x
)dx  } {
a_{3}+b_{3}x   }\nn
\ea
can be written in terms of multiple polylogarithms with the help of the relations
presented  in this paper.
It is worthwhile to mention that all the equations presented in the present 
paper have been also checked numerically.

We have found several examples where the representation of the $L$ functions in 
terms of multiple polylogarithms is not unique. This reflects the fact that 
multiple polylogarithms obey  
quasishuffle and shuffle Hopf algebras and hence satisfy numerous identities
as is the case for the classical polylogarithms.
More information about identities  between multiple polylogarithms can be found, 
e.g., in \cite{nested} and \cite{Vollinga}, and references therein.

For future parton model applications of our results numerical efficiency is 
an important issue. We are presently writing numerical 
$C$\raisebox{0.05cm}{\scriptsize++} 
codes to
compare the numerical efficiency of the two representations in terms of
$L$ functions and multiple polylogarithms.

\vglue 1.0cm
              \begin{bf}
\noindent
ACKNOWLEDGMENTS
               \end{bf}
\vglue .3cm
One of the authors (Z.~M.) would like to thank the Particle Theory Group of the 
Institut f{\"u}r Physik, Universit{\"a}t Mainz, for hospitality.
The work of one of the authors (Z.~M.) was supported by a DFG (Germany) grant under contract No.
436 GEO 17/6/05. One of the authors (M.~R.) was 
supported by the DFG through the Graduiertenkolleg ``Eichtheorien''
at the University of Mainz and by the Helmholtz Gemeinschaft
under contract No. VH-NG-105.

\setcounter{equation}{0}
\renewcommand{\theequation}{A\arabic{equation}}

\vglue 1.0cm               
               \begin{bf}
\noindent 
APPENDIX
               \end{bf}
\vglue .3cm%

In this Appendix we consider as an example the real part of the 
${\cal O}(\ep^2)$ coefficient ${\rm Re}\,D_1^{(2)}$ of the Laurent series 
expansion of the massive box $D_1$ with three massive
propagators. Using the rules written down in the main text of this paper we have 
expressed the corresponding results of \cite{KMR} involving $L$ functions
in terms of multiple polylogarithms. The $L$ function structure of 
${\rm Re}\,D_1^{(2)}$ in \cite{KMR} is sufficiently rich to provide an 
illustration 
of the corresponding complexity in terms of multiple polylogarithms when 
transforming to the latter representation. We mention that all multiple 
polylogarithms up to weight three 
have been reexpressed in terms of classical polylogarithms. We then used
automatic program codes to simplify the classical polylogarithms as much
as possible, as was also done in \cite{KMR}.   
 
We use the notation and the conventions of \cite{KMR}. In brief, 
we use the Mandelstam-type variables
\be
\label{s-t-u}
s\equiv (p_1+p_2)^2, \quad  t\equiv T-m^2
\equiv (p_1-p_3)^2-m^2, 
\quad  u\equiv U-m^2\equiv (p_2-p_3)^2-m^2
\ee
for the $2\to 2$ partonic process 
${a}(p_1)+{b}(p_2)\to {Q}(p_3)+{\overline Q}(p_4)$ with
$p_1^2=p_2^2=0$ and $p_3^2=p_4^2=m^2$. We also introduce the abbreviations
$\Big(\beta=\sqrt{1-4m^2/s}\Big)$
\ba
\label{notations}
\nn  &&
z_3 \equiv (s + 2 t + s\beta)/2,  {\rm \hspace{.4in}}  
z_4 \equiv (s + 2 t - s\beta)/2,             \\
&&    z_5 \equiv (2 m^2 + t + t \beta)/2,  {\rm \hspace{.4in}} 
z_6 \equiv (2 m^2 + t - t \beta)/2,     \\
\nn &&
l_s \equiv \ln \frac{s}{m^2},  {\rm \hspace{.2in}}
l_t \equiv \ln \frac{-t}{m^2},  {\rm \hspace{.2in}}
l_T \equiv \ln \frac{-T}{m^2},  {\rm \hspace{.2in}}
l_x \equiv \ln x,        {\rm \hspace{.2in}}                         \\
\nn &&
l_{\beta}  \equiv \ln \beta,    {\rm \hspace{.2in}}
l_{z3}  \equiv  \ln \frac{z_3}{m^2},     {\rm \hspace{.2in}}
l_{z4}  \equiv  \ln \frac{-z_4}{m^2} \, .
\ea
One finds
\ba
{\rm Re}\,D_1^{(2)}&=& \frac{1}{st\beta} \left[ \frac{1}{192} \left\{ - 109 l_s^4 
     + 240 l_t^4 + 32 l_T^3 l_x + 264 l_T^2 l_x^2 - 200 l_T l_x^3 -
      177 l_x^4 - 
\right.\right.
\\  \nn  &&
     96 l_T^2 l_x l_{z3} + 192 l_T l_x^2 l_{z3} + 12 l_x^3 l_{z3} +
     96 l_T l_x l_{z3}^2 - 32 l_x l_{z3}^3 - 480 l_T^2 l_x l_{z4} + 
\\  \nn  &&
     24 l_T l_x^2 l_{z4} + 180 l_x^3 l_{z4} -
     144 l_x^2 l_{z3} l_{z4} + 480 l_T l_x l_{z4}^2 -
     336 l_x^2 l_{z4}^2 - 96 l_x l_{z3} l_{z4}^2 + 
\\  \nn  &&
     320 l_x l_{z4}^3 - 168 l_{z4}^4 + 480 l_T^2 l_x l_{\beta} - 
     480 l_T l_x^2 l_{\beta} -
     40 l_x^3 l_{\beta} - 192 l_T l_x l_{z3} l_{\beta} +
\\  \nn  &&
     336 l_x^2 l_{z3} l_{\beta} + 96 l_x l_{z3}^2 l_{\beta} -
     384 l_T l_x l_{z4} l_{\beta} + 
     336 l_x^2 l_{z4} l_{\beta} + 192 l_x l_{z3} l_{z4} l_{\beta} + 
\\  \nn  &&
      96 l_x l_{z4}^2 l_{\beta} + 192 l_{z4}^3 l_{\beta} + 
      96 l_T l_x l_{\beta}^2 + 24 l_x^2 l_{\beta}^2 - 
      96 l_x l_{z3} l_{\beta}^2 - 96 l_{z4}^2 l_{\beta}^2 + 
      32 l_x l_{\beta}^3 + 
\\  \nn  &&
      32 l_{\beta}^4 - 32 l_t^3
   ( 9 l_T + 20 l_x + 25 l_{z3} + l_{z4} + 13 l_{\beta}   ) -
     4 l_s^3   (  8 l_T - 36 l_x - 9 l_{z3} - 
\\  \nn  &&
     43 l_{z4} + 94 l_{\beta} ) -
     6 l_s^2 \left( 52 l_t^2 - 28 l_T^2 + 15 l_x^2 + 26 l_x l_{z3} +
       46 l_x l_{z4} + 24 l_{z3} l_{z4} +    \right.
\\  \nn  &&
     32 l_{z4}^2 - 4 l_T
   ( 9 l_x + 8 l_{z3} - 15 l_{z4} - 4 l_{\beta} )
       - 60 l_x l_{\beta} - 24 l_{z3} l_{\beta} -
       56 l_{z4} l_{\beta} + 76 l_{\beta}^2 + 
\\  \nn  && \left.
       l_t (  -44 l_T +  60 l_x - 8 l_{z3} - 60 l_{z4} +
       8 l_{\beta}^{} ) \right) - 24 l_t^2
    \left( 4 l_T^2 + 15 l_x^2 - 8 l_T ( 2 l_x +    \right.
\\  \nn  && 
        2 l_{z3} + 4 l_{z4} - 3 l_{\beta} )
        + 4 l_x ( 3 l_{z3} - 8 l_{z4} + 13 l_{\beta}  ) +
 2 ( -4 l_{z3}^2 + l_{z4}^2 + 12 l_{z3} l_{z4} - 
\\  \nn  &&   \left.
     12 l_{z3} l_{\beta} - 8 l_{z4} l_{\beta} + 10 l_{\beta}^2  ) \right) +
     8 l_t \left( 8 l_T^3 - 31 l_x^3 + l_x^2 ( -6 l_{z3} + 33 l_{z4} 
     + 6 l_{\beta} ) +     \right.
\\  \nn  &&
       6 l_x ( 3 l_{z3}^2 -
         22 l_{z4}^2 - 2 l_{z3} (  5 l_{z4} -
           9 l_{\beta} ) + 20 l_{z4} l_{\beta} - 10 l_{\beta}^2  ) +
       4 ( -2 l_{z3}^3 + 6 l_{z3}^2 l_{z4} +
\\  \nn  &&
         16 l_{z4}^3 + 9 l_{z3}
    ( l_{z4} - l_{\beta} )^2 -
         18 l_{z4}^2 l_{\beta} + 9 l_{z4} l_{\beta}^2 + 2 l_{\beta}^3) +
      6 l_T^2 ( 3 l_x - 4 l_{z3} + 4 l_{z4} - 
\\  \nn  &&  \left.
        4 l_{\beta}  ) - 3 l_T (5 l_x^2 - 4 l_x ( 5 l_{z3} - 4 l_{z4} ) -
      8 ( l_{z3}^2 - 2 l_{z3} l_{z4} -
           3 l_{z4}^2 + 4 l_{z4} l_{\beta} + l_{\beta}^2  )) \right)  -
\\  \nn  &&
      4 l_s \left( 80 l_t^3 + 8 l_T^3 + 18 l_x^3 -
       27 l_x^2 l_{z3} - 8 l_{z3}^3 -
       165 l_x^2 l_{z4} - 72 l_x l_{z3} l_{z4} + 48 l_x 
        l_{z4}^2 -      \right.
\\  \nn  &&
       24 l_{z3} l_{z4}^2 -
       64 l_{z4}^3 + 18 l_x^2 l_{\beta} +
       120 l_x l_{z3} l_{\beta} + 24 l_{z3}^2 l_{\beta} + 
        72 l_x l_{z4} l_{\beta} + 48 l_{z3} l_{z4} l_{\beta} + 
\\  \nn  &&
        72 l_{z4}^2 
        l_{\beta} - 84 l_x l_{\beta}^2 -
       24 l_{z3} l_{\beta}^2 - 48 l_{z4}
        l_{\beta}^2 + 40 l_{\beta}^3 -
       12 l_T^2 (13 l_x + 2 l_{z3} -
         6 l_{z4} + 
\\  \nn  &&
        6 l_{\beta} ) + 12 l_t^2 ( 8 l_T + 5 l_x -
         26 l_{z3} - 10 l_{z4} + 12 l_{\beta}  )
         - 12 l_t
   ( 5 l_T^2 + 6 l_x^2 - 7 l_{z3}^2 -
\\  \nn  &&
         10 l_{z3} l_{z4} - 16 l_{z4}^2 +
         l_x ( 14 l_{z3} + 27 l_{z4} - 4 l_{\beta} )
          + 10 l_{z3} l_{\beta} + 16 l_{z4} l_{\beta} +
         2 l_{\beta}^2 + 
\\  \nn  &&
         l_T ( -5 l_x - 2 l_{z3} + 4 l_{z4} + 8 l_{\beta}  ))
          + 12 l_T
   ( 11 l_x^2 + l_x ( 8 l_{z3} +
           7 l_{z4} - 4 l_{\beta} ) +
      2 l_{z3}^2 - 
\\  \nn  && \left.\left.
           6 l_{z4}^2 - 4 l_{z3} l_{\beta} + 
            8 l_{z4} l_{\beta} + 2 l_{\beta}^2   )\right)\right\} +
   \left( 3 l_s^2/4 - 7 l_t^2/2 + l_T l_x + 3 l_x^2/4 + 5 l_x l_{z3} + \right.
\\  \nn  &&
     5 l_x l_{z4} + l_{z4}^2/2 + 
     l_t ( 2 l_T - l_x + 10 l_{z3} +
       2 l_{z4} - 15 l_{\beta}) -
     8 l_x l_{\beta} + 2 l_{z4} l_{\beta} - 11 l_{\beta}^2 -
\\  \nn  &&  \left.
l_s ( 7 l_t + l_T +
       8 l_x + 5 l_{z3} + l_{z4} +
       6 l_{\beta}^{}  )\right) \zeta(2) +
   ( -3 l_s + 2 l_x) \zeta(3) - 35 \zeta(4)/4 -
\\  \nn  &&
   2 {\rm Li}_2^2\left(\frac{m^2}{z_5}\right) + 
   2 {\rm Li}_2^2\left(\frac{-t (1 - \beta)}{2 m^2}\right) +
   \frac{1}{8} {\rm Li}_2\left(\frac{m^2}{z_5}\right) 
     \left( -11 l_s^2 - 4 l_t^2 + 25 l_x^2 +    \right.
\\  \nn  &&
     8 l_x l_{z4} - 24 l_{z4}^2 + l_s ( 4 l_t +
       26 l_x + 24 l_{z4} - 8 l_{\beta} ) +
     24 l_x l_{\beta} + 16 l_{z4} l_{\beta} - 8 l_{\beta}^2 -
\\  \nn  &&   \left.
     4 l_t ( 9 l_x - 4 l_{z4} +
       4 l_{\beta}^{})\right) + {\rm Li}_2\left(\frac{m^2 x}{-T}\right)
    \left( 11 l_s^2/8 - l_t^2 +
     2 l_t ( l_x + l_{z4} - 2 l_{\beta}) +         \right.
\\  \nn  &&    \left.
l_s ( -l_t +
       7 l_x/4 - l_{z4} + 2 l_{\beta}) +
       l_x ( 19 l_x - 24 l_{z4} +
       16 l_{\beta}^{} )/8 \right) + \frac{1}{8} {\rm Li}_2(-x) \times
\\  \nn  &&
     \left( -15 l_s^2 + 4 l_t^2 + l_x ( -39 l_x + 32 l_{z4} - 
       16 l_{\beta} ) + 2 l_s ( 10 l_t + 5 l_x - 8 l_{\beta} ) +  \right.
\\  \nn  &&    \left.
       l_t ( -44 l_x + 32 l_{\beta}^{} )\right) + 
      {\rm Li}_2\left(\frac{z_3}{z_4}\right) 
      \left( -2 {\rm Li}_2\left(\frac{m^2}{z_5}\right) 
      + \frac{1}{4} \left( -5 l_s^2 + 20 l_t^2 - 7 l_x^2 -  
\right.\right.
\\  \nn  && 
       4 l_x l_{z3} + 12 l_x l_{z4} - 4 l_{z4}^2 +
       l_s ( -2 l_t + 4 ( 2 l_x + l_{z3} + l_{z4} - l_{\beta} )) -
       4 l_x l_{\beta} + 
\\  \nn  &&   \left.
         l_t ( -22 l_x - 8 l_{z3} + 8 l_{\beta}^{} ) - 
         8 \zeta(2) \right) \bigg) + 
   {\rm Li}_2(x)\left( 2 {\rm Li}_2\left(\frac{m^2}{z_5}\right) + 
\right.
\\  \nn  && 
    2 {\rm Li}_2\left(\frac{-t (1 - \beta)}{2 m^2}\right) -
      7 l_s^2/4 - 3 l_t^2/2 -
      11 l_x^2/4 + l_x l_{z3} +
     3 l_x l_{z4} + l_{z4}^2 -
\\  \nn  && 
     2 l_x l_{\beta} + l_t
      ( -4 l_x + 2 l_{z3} + 4 l_{\beta} ) +
       l_s ( 4 l_t + l_x/2 - l_{z3} - l_{z4} - 2 l_{\beta} ) 
       - 6 \zeta(2) \bigg) +
\\  \nn  && 
     {\rm Li}_2\left(\frac{T}{m^2}\right) 
     \left( -4 {\rm Li}_2(x) - 4 {\rm Li}_2\left(\frac{z_3}{z_4}\right) - 
     4 {\rm Li}_2\left(\frac{m^2}{z_5}\right) +
     4 {\rm Li}_2\left(\frac{-t (1 - \beta)}{2 m^2}\right) + \right.
\\  \nn  && 
      l_s^2/8 - 3 l_t^2/2 + l_x^2/8 + 
     2 l_x l_{z3} - 4 l_{z4}^2 + l_t ( 3 l_x/2 +
       4 l_{z3} + 4 l_{z4} - 4 l_{\beta} ) -
\\  \nn  &&   \left.
     2 l_x l_{\beta} + 4 l_{z4} l_{\beta} - 2 l_{\beta}^2 -
      \frac{l_s}{4} ( 18 l_t + 7 l_x +
         8 l_{z3} - 16 l_{z4} + 8 l_{\beta} ) -
       4 \zeta(2) \right) + 
\\  \nn  &&
    {\rm Li}_2\left(\frac{T}{z_3}\right) \left(-\frac{l_s^2}{8} - 2 l_t^2 
     + l_s (\frac{9}{4} l_x - l_{z4}) - 2 l_t (l_x - l_{z4}) -
     l_x (\frac{17}{8} l_x - l_{z4}) \right) +
\\  \nn  &&
      {\rm Li}_2\left(\frac{-t (1 - \beta)}{2 m^2}\right)
    \left( -2 {\rm Li}_2\left(\frac{z_3}{z_4}\right) - 7 l_t^2 - l_x^2/2 +
     2 l_x l_{z3} - l_x l_{z4} - l_{z4}^2 +    \right.
\\  \nn  && 
l_s ( l_t +
       3 l_x/2 - 2 l_{z3} + l_{z4} -
       l_{\beta} ) + 3 l_x l_{\beta} +
     2 l_{z4} l_{\beta} - l_{\beta}^2 -
     l_t ( l_x - 4 l_{z3} -
       2 l_{z4} + 
\\  \nn  && 
       2 l_{\beta}) + 12 \zeta(2) \bigg) + 
   {\rm Li}_3\left(\frac{-1 + \beta}{2\beta}\right) (4 l_s - 7 l_t) +
   5 {\rm Li}_3\left(\frac{z_5}{t\beta}\right) l_t + 
      {\rm Li}_3\left(\frac{m^2}{z_5}\right) \times
\\  \nn  && 
    (5 l_s - 6 l_t - 11 l_x)/2 - 
   {\rm Li}_3\left(\frac{z_3}{t}\right) (4 l_t + 6 l_x) +
    {\rm Li}_3\left(\frac{z_6}{m^2}\right) (3 l_s/2 - 5 l_t - 
\\  \nn  && 
     7 l_x/2 ) + 4 {\rm Li}_3\left(\frac{z_4}{t}\right) (l_s -
     l_t - 2 l_x) + {\rm Li}_3\left(-\frac{m^2 x z_3}{sT\beta}\right)
    (l_s - 2 l_t - l_x) +
\\  \nn  && 
    {\rm Li}_3\left(\frac{-x^2}{1 - x^2}\right) (l_s - l_t -
     l_x) + {\rm Li}_3\left(\frac{z_3}{s\beta}\right) (l_s +
     5 l_t - l_x) + 2 {\rm Li}_3\left(\frac{m^2}{-t}\right) l_x +
\\  \nn  &&
     {\rm Li}_3\left(\frac{z_5}{T}\right) 
    \left( -\frac{5}{2} l_s - 3 l_t - \frac{5}{2} l_x + 4 l_{z4} - 4 
                       l_{\beta}\right) +
      {\rm Li}_3\left(\frac{-t (1 - \beta)}{2 z_5}\right) \times
\\  \nn  &&
    \left( -\frac{5}{2} l_s - l_t - \frac{3}{2} l_x + 4 l_{z4} - 4 
                      l_{\beta}\right) + 
     {\rm Li}_3\left(\frac{-2 z_6}{t(1 + \beta)}\right)
    \left( -\frac{3}{2} l_s - 2 l_t + 2 l_{z4} -     \right.
\\  \nn  &&  \left.
       \frac{l_x}{2} - 2 l_{\beta}\right) +
    {\rm Li}_3\left(\frac{z_3}{z_4}\right) \left( \frac{l_s}{2} - l_t - 
       \frac{l_x}{2} + 2 l_{z4} - 2 l_{\beta}\right) +
   2 {\rm Li}_3(-x) (3 l_s + 
\\  \nn  &&
     2 l_t +  2 l_x - 4 l_{z4} +
     4 l_{\beta}) + {\rm Li}_3\left(\frac{z_6}{z_5}\right)
    \left( \frac{l_s}{2} + 4 l_t - \frac{l_x}{2} - 2 l_{z4} + 2 l_{\beta} 
      \right) +
\\  \nn  &&
   {\rm Li}_3\left(\frac{2 z_6}{m^2 (1 + \beta)}\right) 
    (\frac{3}{2} l_s + 2 l_t + \frac{l_x}{2} - 2 l_{z4} + 2 l_{\beta}) + 
    {\rm Li}_3\left(\frac{z_4}{T}\right) (3 l_s + 3 l_t + 
\\  \nn  &&
     3 l_x - 4 l_{z4} + 4 l_{\beta}) + 
     {\rm Li}_3\left(\frac{T}{z_3}\right) \left( \frac{l_s}{2} +
     l_t + \frac{7}{2} l_x - 2 l_{z4} + 2 l_{\beta} \right) +
\\  \nn  && 
     {\rm Li}_3\left(\frac{m^2 (1 - \beta)}{2 z_5}\right)
    \left( \frac{3}{2} l_s + 3 l_t + \frac{5}{2} l_x - 4 l_{z4} + 
     4 l_{\beta} \right) +
\\  \nn  &&
   {\rm Li}_3(x) \left( \frac{15}{2} l_s - 2 l_t +
     \frac{5}{2} l_x - 6 l_{z4} + 6 l_{\beta} \right) + 
     {\rm Li}_3\left(\frac{T}{z_6}\right) ( - 2 l_s +
\\  \nn  &&
     l_t - 2 l_x + 2 l_{z4} - 2 l_{\beta}) + 
   2 {\rm Li}_4(x) - 4 {\rm Li}_4\left(\frac{z_3}{t}\right) + 
   4 {\rm Li}_4\left(\frac{z_4}{t}\right) -
   {\rm Li}_4\left(\frac{z_4}{T}\right) - 
\\  \nn  &&
   {\rm Li}_4\left(\frac{T z_4}{D}\right) + 
   {\rm Li}_4\left(\frac{z_5}{T}\right) + 
   2 {\rm Li}_4\left(\frac{s(1 - \beta)}{-2t}\right) + 
   3 {\rm Li}_4\left(\frac{s (1 - \beta)}{2 z_4}\right) +
\\  \nn  &&
   {\rm Li}_4\left(\frac{T}{z_6}\right) +
   4 {\rm Li}_4\left(\frac{-1 + \beta}{2\beta}\right) + 
   2 {\rm Li}_4\left( \frac{-2t}{s(1 + \beta)}\right) +
   4 {\rm Li}_4\left(\frac{2\beta}{1 + \beta}\right) + 
\\  \nn  &&
   3 {\rm Li}_4\left(\frac{2 z_3}{s(1 + \beta)}\right) +
   2 Li_{3, 1} \left(-\frac{m^2 x z_3}{s T \beta}, \frac{-T}{m^2 x}\right) -
\\  \nn  &&
   2 Li_{3, 1} \left(-\frac{m^2 x z_3}{s T \beta}, 
                                            \frac{2T}{t(1-\beta)}\right) -
   6 Li_{1, 2, 1} \left(1, \frac{s (1 - \beta)}{2 z_4}, 
                                                  \frac{z_5}{m^2}\right) +
\\  \nn  &&
   6 Li_{1, 2, 1} \left(1, \frac{s(1 + \beta)}{2 z_3}, 
                                                  \frac{z_6}{m^2}\right) -
   2 Li_{2, 1, 1} \left(1, \frac{z_4}{z_3}, \frac{z_3}{t}\right) -
\\  \nn  &&
   2 Li_{2, 1, 1} \left(\frac{m^2}{T}, \frac{T}{z_5}, 
                                                  \frac{z_5}{m^2}\right) +  
   2 Li_{2, 1, 1} \left(\frac{m^2}{T}, \frac{T}{z_6}, 
                                                  \frac{z_6}{m^2}\right) -
\\  \nn  &&
   2 Li_{2, 1, 1} \left(\frac{z_3}{z_4}, \frac{z_4}{z_3}, 
                                                    \frac{z_3}{t}\right) - 
   2 Li_{2, 1, 1} \left(\frac{m^2}{z_5}, 1, \frac{z_5}{m^2}\right) - 
   2 Li_{2, 1, 1} \left(\frac{m^2}{z_5}, \frac{z_5}{T}, 
                                                    \frac{T}{m^2}\right) +
\\  \nn  &&
   2 Li_{2, 1, 1} \left(\frac{s(1 - \beta)}{2 z_4}, \frac{z_5}{z_6}, 
                                                  \frac{z_6}{m^2}\right) +
   2 Li_{2, 1, 1} \left(-\frac{m^2 x z_3}{sT\beta}, -\frac{sT\beta}{m^2xz_3},
                                               \frac{z_3}{s\beta}\right) - 
\\  \nn  &&
   2 Li_{2, 1, 1} \left(-\frac{m^2xz_3}{sT\beta},
                    -\frac{sT\beta}{m^2xz_3}, -\frac{z_6}{t\beta}\right) +
   2 Li_{2, 1, 1} \left(\frac{s(1 + \beta)}{2z_3}, 1, 
                                                  \frac{z_6}{m^2}\right) +
\\  \nn  &&
   2 Li_{2, 1, 1} \left(\frac{s(1 + \beta)}{2z_3}, \frac{z_6}{T}, 
                                                    \frac{T}{m^2}\right) -
   2 Li_{2, 1, 1} \left(\frac{s(1 + \beta)}{2z_3}, \frac{z_6}{z_5}, 
                                                  \frac{z_5}{m^2}\right) +
\\  \nn  &&
   Li_{1, 1, 1, 1} \left(1, \frac{T}{z_6}, \frac{z_6}{z_5}, 
                                                  \frac{z_5}{m^2}\right) -
   Li_{1, 1, 1, 1} \left(1, \frac{s(1 - \beta)}{2z_4}, \frac{z_5}{T}, 
                                                    \frac{T}{m^2}\right) +
\\  \nn  &&
   3 Li_{1, 1, 1, 1} \left(1, \frac{s(1 - \beta)}{2z_4}, \frac{z_5}{z_6}, 
                                                  \frac{z_6}{m^2}\right) +
   Li_{1, 1, 1, 1} \left(1, \frac{s(1 + \beta)}{2z_3}, \frac{z_6}{T}, 
                                                    \frac{T}{m^2}\right) -
\\  \nn  &&
   3 Li_{1, 1, 1, 1} \left(1, \frac{s(1 + \beta)}{2z_3}, \frac{z_6}{z_5}, 
                                                  \frac{z_5}{m^2}\right) -
   2 Li_{1, 1, 1, 1} \left(\frac{t}{T}, 1, \frac{T}{z_3}, 
                                                    \frac{z_3}{t}\right) + 
\\  \nn  &&
   2 Li_{1, 1, 1, 1} \left(\frac{t}{T}, 1, \frac{T}{z_4}, 
                                                    \frac{z_4}{t}\right) +
   2 Li_{1, 1, 1, 1} \left(\frac{t}{T}, \frac{T}{z_3}, 1, 
                                                    \frac{z_3}{t}\right) -
\\  \nn  &&
   2 Li_{1, 1, 1, 1} \left(\frac{t}{T}, \frac{T}{z_3}, \frac{z_3}{T}, 
                                                      \frac{T}{t}\right) -
   2 Li_{1, 1, 1, 1} \left(\frac{t}{T}, \frac{T}{z_3}, \frac{z_3}{z_4}, 
                                                    \frac{z_4}{t}\right) -
\\  \nn  &&
   2 Li_{1, 1, 1, 1} \left(\frac{t}{T}, \frac{T}{z_4}, 1, 
                                                    \frac{z_4}{t}\right) +
   2 Li_{1, 1, 1, 1} \left(\frac{t}{T}, \frac{T}{z_4}, \frac{z_4}{T}, 
                                                      \frac{T}{t}\right) +  
\\  \nn  &&
   2 Li_{1, 1, 1, 1} \left(\frac{t}{T}, \frac{T}{z_4}, \frac{z_4}{z_3}, 
                                                    \frac{z_3}{t}\right) +
   2 Li_{1, 1, 1, 1} \left(\frac{t}{z_3}, 1, \frac{z_3}{T}, 
                                                      \frac{T}{t}\right) -
\\  \nn  &&
   2 Li_{1, 1, 1, 1} \left(\frac{t}{z_3}, \frac{z_3}{T}, 1, 
                                                      \frac{T}{t}\right) +
   2 Li_{1, 1, 1, 1} \left(\frac{t}{z_3}, \frac{z_3}{T}, \frac{T}{z_3}, 
                                                    \frac{z_3}{t}\right) -
\\  \nn  &&
   2 Li_{1, 1, 1, 1} \left(\frac{t}{z_3}, \frac{z_3}{T}, \frac{T}{z_4}, 
                                                    \frac{z_4}{t}\right) -
   2 Li_{1, 1, 1, 1} \left(\frac{t}{z_3}, \frac{z_3}{z_4}, \frac{z_4}{T}, 
                                                      \frac{T}{t}\right) -
\\  \nn  &&
   2 Li_{1, 1, 1, 1} \left(\frac{t}{z_4}, 1, \frac{z_4}{T}, 
                                                      \frac{T}{t}\right) +
   2 Li_{1, 1, 1, 1} \left(\frac{t}{z_4}, \frac{z_4}{T}, 1, 
                                                      \frac{T}{t}\right) +
\\  \nn  &&
   2 Li_{1, 1, 1, 1} \left(\frac{t}{z_4}, \frac{z_4}{T}, \frac{T}{z_3}, 
                                                    \frac{z_3}{t}\right) -
   2 Li_{1, 1, 1, 1} \left(\frac{t}{z_4}, \frac{z_4}{T}, \frac{T}{z_4}, 
                                                    \frac{z_4}{t}\right) +
\\  \nn  &&
   2 Li_{1, 1, 1, 1} \left(\frac{t}{z_4}, \frac{z_4}{z_3}, \frac{z_3}{T}, 
                                                      \frac{T}{t}\right) -
   2 Li_{1, 1, 1, 1} \left(\frac{T}{z_4}, \frac{z_4}{z_3}, 1, 
                                                    \frac{z_3}{t}\right) +
\\  \nn  &&
   Li_{1, 1, 1, 1} \left(\frac{z_4}{T}, \frac{T}{z_3}, 1, 
                                                    \frac{z_3}{t}\right) - 
   Li_{1, 1, 1, 1} \left(\frac{m^2}{z_5}, 1, \frac{z_5}{T}, 
                                                    \frac{T}{m^2}\right) + 
\\  \nn  &&
   Li_{1, 1, 1, 1} \left(\frac{m^2}{z_5}, \frac{z_5}{T}, 1, 
                                                    \frac{T}{m^2}\right) -
   Li_{1, 1, 1, 1} \left(\frac{m^2}{z_5}, \frac{z_5}{T}, \frac{T}{z_5}, 
                                                  \frac{z_5}{m^2}\right) -
\\  \nn  &&
   Li_{1, 1, 1, 1} \left(\frac{m^2}{z_6}, \frac{z_6}{T}, 1, 
                                                    \frac{T}{m^2}\right) -
   Li_{1, 1, 1, 1} \left(\frac{m^2}{z_6}, \frac{z_6}{T}, \frac{T}{z_5}, 
                                                  \frac{z_5}{m^2}\right) -
\\  \nn  &&
   Li_{1, 1, 1, 1} \left(\frac{m^2}{z_6}, \frac{z_6}{z_5}, 1, 
                                                  \frac{z_5}{m^2}\right) -  
   Li_{1, 1, 1, 1} \left(\frac{m^2}{z_6}, \frac{z_6}{z_5}, \frac{z_5}{T}, 
                                                    \frac{T}{m^2}\right) -
\\  \nn  && 
   Li_{1, 1, 1, 1} \left(\frac{z_6}{T}, \frac{T}{z_5}, 1, 
                                                  \frac{z_5}{m^2}\right) +
   Li_{1, 1, 1, 1} \left(\frac{z_6}{z_5}, \frac{z_5}{T}, 1, 
                                                    \frac{T}{m^2}\right) -
\\  \nn  && 
   3 Li_{1, 1, 1, 1} \left(\frac{s(1 - \beta)}{2z_4}, 1, \frac{z_5}{z_6}, 
                                                  \frac{z_6}{m^2}\right) +
   Li_{1, 1, 1, 1} \left(\frac{s(1 - \beta)}{2z_4}, \frac{z_5}{T}, 
                                   \frac{T}{z_6}, \frac{z_6}{m^2}\right) -
\\  \nn  &&
   3 Li_{1, 1, 1, 1} \left(\frac{s(1 - \beta)}{2z_4}, \frac{z_5}{z_6}, 1, 
                                                  \frac{z_6}{m^2}\right) +
   Li_{1, 1, 1, 1} \left(\frac{s(1 - \beta)}{2z_4}, \frac{z_5}{z_6}, 
                                     \frac{z_6}{T}, \frac{T}{m^2}\right) -
\\  \nn  &&
   3 Li_{1, 1, 1, 1} \left(\frac{s(1 - \beta)}{2z_4}, \frac{z_5}{z_6}, 
                                 \frac{z_6}{z_5}, \frac{z_5}{m^2}\right) +
   Li_{1, 1, 1, 1} \left(\frac{s(1 + \beta)}{2z_3}, 1, \frac{z_6}{T}, 
                                                    \frac{T}{m^2}\right) - 
\\  \nn  &&
   Li_{1, 1, 1, 1} \left(\frac{s(1 + \beta)}{2z_3}, 1, \frac{z_6}{z_5}, 
                                                  \frac{z_5}{m^2}\right) +
   Li_{1, 1, 1, 1} \left(\frac{s(1 + \beta)}{2z_3}, \frac{z_6}{T}, 
                                   \frac{T}{z_6}, \frac{z_6}{m^2}\right) -
\\  \nn  &&
   Li_{1, 1, 1, 1} \left(\frac{s(1 + \beta)}{2z_3}, \frac{z_6}{z_5}, 
                                \frac{z_5}{z_6}, \frac{z_6}{m^2}\right) -
   i \pi (l_s - 2 l_t - l_x)
    \left( \frac{l_s^2}{2} + l_s l_T + l_s l_x +   \right.
\\  \nn  &&
     \frac{l_x^2}{2} - l_T l_{z3} + \frac{l_{z3}^2}{2} -
     l_s l_{z4} + \frac{l_{z4}^2}{2} + l_s l_{\beta} + l_T
      l_{\beta} + l_x l_{\beta} - l_{z4} l_{\beta} + \frac{l_{\beta}^2}{2} +
\\  \nn  &&  \left.\left.
     2 {\rm Li}_2(-x) + 2 {\rm Li}_2(x) - 
     {\rm Li}_2\left(\frac{m^2 x}{-T}\right) 
    + {\rm Li}_2\left(\frac{T}{z_3}\right) +  
     {\rm Li}_2\left(\frac{z_3}{z_4}\right) + \zeta(2) \right) 
\right].
\ea
At the very end of the expression one finds an explicit imaginary part. Since
the whole expression must be real this clearly indicates that the same imaginary 
contribution with opposite sign must be contained in multiple polylogarithms, 
e.g., some of them are sitting on branch cuts. This is in fact true for the
multiple polylogarithms 
\ba  \nn  &&
Li_{3, 1} \left(-\frac{m^2 x z_3}{s T \beta}, \frac{-T}{m^2 x}\right), 
\qquad
Li_{3, 1} \left(-\frac{m^2 x z_3}{s T \beta}, \frac{2T}{t(1-\beta)}\right), 
\\ \nn   &&
Li_{2, 1, 1} \left(-\frac{m^2 x z_3}{sT\beta}, -\frac{sT\beta}{m^2xz_3},
                                      \frac{z_3}{s\beta}\right), \qquad
Li_{2, 1, 1} \left(-\frac{m^2xz_3}{sT\beta}, -\frac{sT\beta}{m^2xz_3},
                                               -\frac{z_6}{t\beta}\right). 
\ea
Indeed, one finds that the imaginary contributions cancel out when one 
numerically evaluates the result.

As regards the length the representations of ${\rm Re}\,D_1^{(2)}$ in terms of
$L$ functions in \cite{KMR} and in terms of multiple polylogarithms are of
similar size. The representation in terms of $L$ functions contains 43 different
$L$ function expressions against 59 different multiple polylogarithm
expressions.

\end{document}